\documentclass[11pt]{article}
\pdfoutput=1
\usepackage{fullpage}
\usepackage{xcolor}
\usepackage{cite}
\usepackage{graphicx}
\usepackage{amsmath}
\usepackage{amssymb}
\usepackage{subcaption}
\usepackage[utf8x]{inputenc} 
\title{}
\date{}

\def\beq{\begin{equation}}
\def\eeq{\end{equation}}
\allowdisplaybreaks
\begin{document}
\bibliographystyle{utphys}

\newcommand{\be}{\begin{equation}}
\newcommand{\ee}{\end{equation}}
\newcommand\n[1]{\textcolor{red}{(#1)}} 
\newcommand{\diff}{\mathop{}\!\mathrm{d}}
\newcommand{\lb}{\left}
\newcommand{\rb}{\right}
\newcommand{\f}{\frac}
\newcommand{\pd}{\partial}
\newcommand{\tr}{\text{tr}}
\newcommand{\fdiff}{\mathcal{D}}
\newcommand{\im}{\text{im}}
\let\caron\v
\renewcommand{\v}{\mathbf}
\newcommand{\T}{\tensor}
\newcommand{\R}{\mathbb{R}}
\newcommand{\C}{\mathbb{C}}
\newcommand{\Z}{\mathbb{Z}}
\newcommand{\msbar}{\ensuremath{\overline{\text{MS}}}}
\newcommand{\DIS}{\ensuremath{\text{DIS}}}
\newcommand{\abar}{\ensuremath{\bar{\alpha}_S}}
\newcommand{\bb}{\ensuremath{\bar{\beta}_0}}
\newcommand{\rc}{\ensuremath{r_{\text{cut}}}}
\newcommand{\Nd}{\ensuremath{N_{\text{d.o.f.}}}}
\newcommand{\red}[1]{{\color{red} #1}}
\setlength{\parindent}{0pt}

\titlepage
\begin{flushright}
QMUL-PH-21-16\\
\end{flushright}

\vspace*{0.5cm}

\begin{center}
{\bf \Large The Weyl double copy from twistor space}

\vspace*{1cm} 
\textsc{Erick Chac\'{o}n\footnote{e.c.chaconramirez@qmul.ac.uk},
Silvia Nagy\footnote{s.nagy@qmul.ac.uk},
  and Chris D. White\footnote{christopher.white@qmul.ac.uk}} \\

\vspace*{0.5cm} Centre for Research in String Theory, School of
Physics and Astronomy, \\
Queen Mary University of London, 327 Mile End
Road, London E1 4NS, UK\\

\end{center}

\vspace*{0.5cm}

\begin{abstract}
The Weyl double copy is a procedure for relating exact solutions in
biadjoint scalar, gauge and gravity theories, and relates fields in
spacetime directly. Where this procedure comes from, and how general
it is, have until recently remained mysterious. In this paper, we show
how the current form and scope of the Weyl double copy can be derived
from a certain procedure in twistor space. The new formalism shows
that the Weyl double copy is more general than previously thought,
applying in particular to gravity solutions with arbitrary Petrov
types. We comment on how to obtain anti-self-dual as well as self-dual
fields, and clarify some conceptual issues in the twistor approach.
\end{abstract}

\vspace*{0.5cm}

\section{Introduction}
\label{sec:intro}

The double copy is by now a highly-studied procedure for turning
solutions in a (non-)abelian gauge theory into gravitational
counterparts. It originally arose in the study of perturbative
scattering amplitudes~\cite{Bern:2010ue,Bern:2010yg}, where it had a
string theoretic motivation~\cite{Kawai:1985xq} at
tree-level. Although the double copy remains conjectural, there is by
now a huge amount of evidence for its being exact in a variety of
(supersymmetric) theories, including at the quantum level, and to all
orders in perturbation theory in certain kinematic
limits~\cite{Bern:2010ue,Bern:1998ug,Green:1982sw,Bern:1997nh,Carrasco:2011mn,Carrasco:2012ca,Mafra:2012kh,Boels:2013bi,Bjerrum-Bohr:2013iza,Bern:2013yya,Bern:2013qca,Nohle:2013bfa,Bern:2013uka,Naculich:2013xa,Du:2014uua,Mafra:2014gja,Bern:2014sna,Mafra:2015mja,He:2015wgf,Bern:2015ooa,Mogull:2015adi,Chiodaroli:2015rdg,Bern:2017ucb,Johansson:2015oia,Oxburgh:2012zr,White:2011yy,Melville:2013qca,Luna:2016idw,Saotome:2012vy,Vera:2012ds,Johansson:2013nsa,Johansson:2013aca,Bargheer:2012gv,Huang:2012wr,Chen:2013fya,Chiodaroli:2013upa,Johansson:2014zca,Johansson:2017srf,Chiodaroli:2017ehv,Chen:2019ywi,Cheung:2020uts}. A
related body of work has extended the double copy to classical
solutions. The first work to appear was the {\it Kerr-Schild double
  copy} of ref.~\cite{Monteiro:2014cda}, which concerned certain exact
(albeit algebraically special) solutions of the Einstein equations,
and demonstrated the existence of well-defined counterparts in gauge
and biadjoint scalar theory. A second exact procedure is the {\it Weyl double
  copy} of ref.~\cite{Luna:2018dpt}, which is more general than the
Kerr-Schild approach, although equivalent to the latter where they
overlap. However, it is still apparently restricted to algebraically
special solutions, albeit ones that may be interesting for
astrophysical purposes (see
e.g. refs.~\cite{Luna:2015paa,Luna:2016due,Carrillo-Gonzalez:2017iyj,Bahjat-Abbas:2017htu,Berman:2018hwd,Bah:2019sda,CarrilloGonzalez:2019gof,Banerjee:2019saj,Ilderton:2018lsf,Monteiro:2018xev,Lee:2018gxc,Cho:2019ype,Kim:2019jwm,Alfonsi:2020lub,Bahjat-Abbas:2020cyb,White:2016jzc,DeSmet:2017rve,Bahjat-Abbas:2018vgo,Elor:2020nqe,Gumus:2020hbb,Keeler:2020rcv,Arkani-Hamed:2019ymq,Huang:2019cja,Alawadhi:2019urr,Moynihan:2019bor,Alawadhi:2020jrv,Easson:2020esh,Casali:2020vuy,Cristofoli:2020hnk,Casali:2020uvr,Pasterski:2020pdk,Adamo:2020qru,Alkac:2021bav,Monteiro:2020plf,Guevara:2020xjx,Momeni:2020hmc}
for follow-up studies). A key feature of exact classical double copies
is that the equations of motion in each theory turn out to be
linearised, with no non-linear corrections. Other classical double
copy techniques have been developed that can in principle go beyond
this~\cite{Luna:2016hge,Goldberger:2016iau,Goldberger:2017frp,Goldberger:2017vcg,Goldberger:2017ogt,Shen:2018ebu,Carrillo-Gonzalez:2018pjk,Plefka:2018dpa,Plefka:2019hmz,Goldberger:2019xef,PV:2019uuv,Anastasiou:2014qba,Borsten:2015pla,Anastasiou:2016csv,Cardoso:2016ngt,Borsten:2017jpt,Borsten:2013bp,Anastasiou:2018rdx,LopesCardoso:2018xes,Luna:2020adi,Borsten:2020xbt,Borsten:2020zgj,Luna:2017dtq,Kosower:2018adc,Maybee:2019jus,Bautista:2019evw,Bautista:2019tdr,Cheung:2018wkq,Bern:2019crd,Bern:2019nnu,Bern:2020buy,Kalin:2019rwq,Kalin:2020mvi,Almeida:2020mrg,Campiglia:2021srh,Borsten:2021,Borsten:2019prq,Borsten:2021hua},
but at the price of proceeding order-by-order in perturbation theory,
such that one loses an exact understanding. One might thus hope that
further scrutiny of the Kerr-Schild and Weyl double copies - including
ascertaining their limitations and scope - will yield an underlying
explanation of where the double copy comes from, and an understanding
of how to apply it to arbitrary solutions. Given the fact that the
Weyl double copy is more general than the Kerr-Schild double copy, it
is sufficient to focus on the former.\\

Recently, a derivation of the Weyl double copy has been
given~\cite{White:2020sfn}, using ideas from twistor
theory~\cite{Penrose:1967wn,Penrose:1968me,Penrose:1972ia} (see
e.g. refs.~\cite{Penrose:1987uia,Penrose:1986ca,Huggett:1986fs,Adamo:2017qyl}
for pedagogical reviews). This explained why the Weyl procedure has
the form it has, as well as suggesting that it is more general than
previously thought. In particular, the original Weyl double copy of
ref.~\cite{Luna:2018dpt} (see also ref.~\cite{Godazgar:2020zbv})
applies to gravitational solutions of Petrov types D and N
only~\footnote{We review the Petrov classification below, after
  introducing the appropriate language.}. However,
ref.~\cite{White:2020sfn} found an example outside these classes, and
also argued that the Weyl double copy should extend to general
conformally flat spacetimes, thus formalising preliminary remarks in
this regard~\cite{Bahjat-Abbas:2017htu,Carrillo-Gonzalez:2017iyj} (see
also~\cite{Alkac:2021bav} for a recent discussion of the classical
double copy in curved space). The aim of the present paper is then
twofold. Firstly, we will fill in the details of the brief
ref.~\cite{White:2020sfn}, providing full details of how to carry out
the appropriate calculations. Secondly, we will extend previous
results to anti-self-dual as well as self-dual fields, as well as to
arbitrary Petrov types.\\

The structure of our paper is as follows. In section~\ref{sec:review},
we will introduce salient details regarding spinors and twistors, as
well as reviewing the Weyl double copy of ref.~\cite{Luna:2018dpt}. In
section~\ref{sec:Weyl}, we will present our twistor-space formalism
for obtaining spacetime Weyl double copy formulae, going beyond the
preliminary results of ref.~\cite{White:2020sfn}. We summarise our
results and conclude in section~\ref{sec:conclude}.

\section{From spinors to twistors}
\label{sec:review}
The Weyl double copy relies on the spinorial formalism of General
Relativity and related theories. Although this is textbook material
(see e.g. ref.~\cite{Stewart:1990uf} in addition to the above
references), this formalism is not necessarily known to all
researchers working on the double copy or beyond, and the same can
certainly be said about twistor theory. We will thus review key
concepts in this section, in order to make our presentation
self-contained, and also to set up crucial notation needed for the
rest of the paper.

\subsection{The spinorial formalism}
\label{sec:spinorial}

Our first introduction to GR typically uses the language of tensors
and four-vectors. However, an alternative formulation exists, in which
all equations are expressed in terms of two-component spinors
$\pi^A\equiv(\pi^0,\pi^1)$, and their higher-rank
generalisations. Spinor indices may be raised and lowered according to
\begin{equation}
\pi_A=\epsilon_{AB}\pi^B,\quad \pi^B=\pi_A\epsilon^{AB},
\label{raiselower}
\end{equation}
where $\epsilon_{AB}$ and $\epsilon^{AB}$ are the two-dimensional
Levi-Civita symbols defined such that~\footnote{Our conventions follow
  those of refs.~\cite{Penrose:1987uia,Penrose:1986ca}, where we have
  chosen an appropriate spin basis to define $\epsilon_{01}$.}
\begin{equation}
\epsilon_{AB}\epsilon^{CB}=\delta^C_A,\quad \epsilon_{01}=1,
\label{Levi-Civita}
\end{equation}
where $\delta^{A}_C$ denotes the Kronecker symbol. Given a spinor
$\pi^A$, one may also consider its complex conjugate $\pi^{A'}$, where
primed indices may be raised and lowered analogously to
eq.~(\ref{raiselower}), but with the symbols $\epsilon^{A'B'}$ and
$\epsilon_{A'B'}$, such that
\begin{equation}
\epsilon_{A'B'}\epsilon^{C'B'}=\delta^{C'}_{A'},\quad \epsilon_{0'1'}=1.
\label{Levi-Civita2}
\end{equation}
These operations extend to objects with any number of spinor
indices. Furthermore, there is a remarkable simplification of the
structure of these higher-rank quantities, which ultimately follows
from the fact that there are only two values that can be held by each
spinor index $A$ or $A'$. Let us introduce the notation
\begin{equation}
\phi_{(A_1A_2\ldots A_n)}=\frac{1}{n!}\sum_{\sigma}
\phi_{A_{\sigma(1)}A_{\sigma(2)}\ldots A_{\sigma(n)}},
\label{symmetrise}
\end{equation}
where the sum is over all permutations $\sigma$ of the index labels
$(1,\ldots,n)$.  That is, eq.~(\ref{symmetrise}) constitutes the fully
symmetric combination of spinor components of an arbitrary rank $n$
spinor (with suitable generalisation to primed indices). Then any
multi-rank spinor can be decomposed into a sum of terms, each of which
involves symmetric spinors, multiplying Levi-Civita symbols. We will
see explicit examples of this shortly. Another nice property is that
any symmetric spinor factorises into a symmetrised product of spinors
e.g.
\begin{equation}
S_{AB\ldots C}=S_{(AB\ldots C)}\quad\Rightarrow\quad
S_{AB\ldots C}=\alpha_{(A}\beta_{B}\ldots \gamma_{C)}.
\label{Sfac}
\end{equation}
The individual spinors $\{\alpha_A,\beta_B,\ldots\}$ are associated
with null vectors in spacetime, referred to as {\it principal null
  directions} of $S_{AB\ldots C}$. To see this, one may note that any
tensorial quantity can be translated into the spinorial language using
the so-called {\it Infeld-van der Waerden symbols}
$\{\sigma^\mu_{AA'}\}$, which may be chosen in Cartesian coordinates
as follows:
\begin{align}
\sigma^0_{AA'}&=\frac{1}{\sqrt{2}}
\left(\begin{array}{rr}1 & 0\\0 & 1\end{array}\right)=\sigma_0^{AA'},\quad 
\sigma^1_{AA'}=\frac{1}{\sqrt{2}}
\left(\begin{array}{rr}0 & 1\\1 & 0\end{array}\right)=\sigma_1^{AA'},\quad\notag\\
\sigma^2_{AA'}&=\frac{1}{\sqrt{2}}
\left(\begin{array}{rr}0 & -i\\i & 0\end{array}\right)=-\sigma_2^{AA'},\quad 
\sigma^3_{AA'}=\frac{1}{\sqrt{2}}
\left(\begin{array}{rr}1 & 0\\0 & -1\end{array}\right)=\sigma_3^{AA'}.
\label{Infeld}
\end{align}
For a 4-vector this gives
\begin{equation}
V_\alpha \sigma^\alpha_{AA'}=
\frac{1}{\sqrt{2}}\left(\begin{array}{cc}
V_0+V_3 & V_1-iV_2\\ V_1+iV_2 & V_0-V_3
\end{array}\right),
\label{4-vector}
\end{equation}
where the determinant of the matrix on the right-hand side is
\begin{equation}
{\rm det}\left(V_\alpha \sigma^\alpha_{AA'}\right)
=\frac12\left((V_0)^2-(V_1)^2-(V_2)^2-(V_3)^2\right).
\label{4-vector2}
\end{equation}
We may recognise this as being proportional to the norm of the
4-vector, such that the determinant vanishes if $V_\alpha$ is null. By
standard linear algebra arguments, this then implies that the matrix
must factorise i.e.
\begin{equation}
V_\alpha V^\alpha=0\quad\Rightarrow\quad 
V_\alpha \sigma^\alpha_{AA'}=\pi_A\pi_{A'},
\label{Vfac}
\end{equation}
where $\pi_{A'}=(\pi_A)^*$ given that the matrix in
eq.~(\ref{4-vector}) is clearly Hermitian. Conversely, given any
spinor $\pi_A$, we may construct a matrix $M_{AA'}=\pi_A \pi_{A'}$,
which in turn corresponds to a null 4-vector in spacetime. In
particular, each of the so-called {\it principal spinors} appearing in
the decomposition of a general symmetric tensor (eq.~(\ref{Sfac})) can
be associated with a {\it principal null direction} in spacetime. \\

A given solution to Einstein's field equations in GR will have a
corresponding Riemann tensor $R_{\alpha\beta\gamma\delta}$. One may
translate this into the spinor language as above, and then decompose
it into various symmetrised spinor parts, where some further
simplifications arise due to the known symmetries of the Riemann
tensor itself. The result turns out to be
\begin{align}
R_{\alpha\beta\gamma\delta}\rightarrow R_{AA'BB'CC'DD'}&=
\Psi_{ABCD}\epsilon_{A'B'}\epsilon_{C'D'}+
\bar{\Psi}_{A'B'C'D'}\epsilon_{AB}\epsilon_{CD}\notag\\
&+\Phi_{ABC'D'}\epsilon_{A'B'}\epsilon_{CD}+
\bar{\Phi}_{A'B'CD}\epsilon_{AB}\epsilon_{C'D'}\notag\\
&+2\Lambda(\epsilon_{AC}\epsilon_{BD}\epsilon_{A'B'}\epsilon_{C'D'}
+\epsilon_{AB}\epsilon_{CD}\epsilon_{A'D'}\epsilon_{B'C'}),
\label{Rdecomp}
\end{align}
where all spinors appearing on the right-hand side are fully
symmetric. The spinors $\Phi_{ABA'B'}$ and $\bar{\Phi}_{A'B'AB}$ are
directly related to the trace-reversed Ricci tensor $R_{\alpha\beta}$,
and $\Lambda$ is directly proportional to the Ricci scalar $R$ (see
e.g. ref.~\cite{Huggett:1986fs}). We will be concerned with vacuum
spacetimes, so that these quantities all vanish by the Einstein
equations. We are then left with the only contribution to the
curvature that is present in free space, which in the usual tensorial
formulation of GR is known as the {\it Weyl tensor}, and denoted
$C_{\alpha\beta\gamma\delta}$. From eq.~(\ref{Rdecomp}), we thus have
the spinorial identification
\begin{align}
C_{\alpha\beta\gamma\delta}\rightarrow 
\Psi_{ABCD}\epsilon_{A'B'}\epsilon_{C'D'}+
\bar{\Psi}_{A'B'C'D'}\epsilon_{AB}\epsilon_{CD}.
\label{Cdecomp}
\end{align}
If we are working in Lorentzian signature such that the spacetime is
real, $\bar{\Psi}_{A'B'C'D'}$ must simply be the complex conjugate of
$\Psi_{ABCD}$. More generally, $\Psi_{ABCD}$ and
$\bar{\Psi}_{A'B'C'D'}$ are the {\it anti-self-dual} and {\it
  self-dual} parts of the Weyl tensor respectively. That is, they are
respectively projected out by the conditions
\begin{equation}
{^*}C_{\alpha\beta\gamma\delta}=\mp i C_{\alpha\beta\gamma\delta},
\label{SDdef}
\end{equation}
where the {\it dual Weyl tensor} is defined by
\begin{equation}
{^*}C_{\alpha\beta\gamma\delta}=\frac12 \epsilon_{\alpha\beta\sigma\tau}
{C^{\sigma\tau}}_{\gamma\delta}.
\label{SDdef2}
\end{equation}
The dynamics of the Weyl tensor is constrained by the Bianchi identity
for the Riemann tensor, which can be shown to lead to the following
conditions:
\begin{equation}
\nabla^{AA'}\Psi_{ABCD}=0,\quad \nabla^{AA'}\bar{\Psi}_{A'B'C'D'}=0,
\label{Einstein}
\end{equation}
where $\nabla^{AA'}=\nabla^\mu \sigma_\mu^{AA'}$ is the appropriate
spinorial translation of the spacetime covariant derivative. Given its
role as part of the Weyl tensor, the spinor $\Psi_{ABCD}$ is usually
referred to as the {\it Weyl spinor}.\\

Let us now turn to electromagnetism, whose equation of motion involves
the field strength tensor $F_{\mu\nu}$. Using similar methods to those
above, one may show that the spinorial decomposition of the field
strength is as follows:
\begin{align}
F_{\alpha\beta}\rightarrow F_{AA'BB'}=\phi_{AB}\epsilon_{A'B'}
+\bar{\phi}_{A'B'}\epsilon_{AB},
\label{Fmunu}
\end{align}
where the symmetric spinors $\phi_{AB}$ and $\bar{\phi}_{A'B'}$ are
the anti-self-dual and self-dual parts, respectively projected out by
\begin{equation}
{^*}F_{\alpha\beta}=\mp iF_{\alpha\beta},\quad
{^*}F_{\alpha\beta}=\frac12 \epsilon_{\alpha\beta\sigma\tau}
F^{\sigma\tau}.
\label{Fdual}
\end{equation}
The Maxwell equations then imply
\begin{equation}
\nabla^{AA'}\phi_{AB}=0,\quad \nabla^{AA'}\bar{\phi}_{A'B'}=0.
\label{Maxwell}
\end{equation}
We may note that eqs.~(\ref{Einstein}) and~(\ref{Maxwell}) are both
special cases of the general spinorial equations (see
e.g. ref.~\cite{Penrose:1987uia})
\begin{equation}
\nabla^{AA'}\phi_{AB\ldots C}=0,\quad \nabla^{AA'}\bar{\phi}_{A'B'\ldots C'}=0
\label{massless}
\end{equation}
where $\phi_{AB\ldots C}$ is assumed symmetric, with $n$
indices. These are known as the {\it massless free field equations},
as they are indeed associated with massless and non-interacting fields
in a vacuum spacetime. The spin of the field is given by the number of
spinor indices divided by two, which matches the two cases described
above: eqs.~(\ref{Einstein}) and~(\ref{Maxwell}) contain spinors with
four and two spinor indices, describing the spin-2 graviton and spin-1
photon respectively. If we restrict to solutions of positive
frequency, the spinors $\bar{\phi}_{A'B'\ldots C'}$ and
$\phi_{AB\ldots C}$ represent states of positive and negative helicity
$\pm n/2$ respectively (in units of $\hbar$)~\footnote{In optics
  parlance, positive and negative helicity correspond to right-handed
  and left-handed circular polarisations.}. Note that we have not yet
stated which spacetime we are working with, which above corresponds to
the fact that $\nabla^{AA'}$ is the covariant derivative associated
with a potentially curved spacetime. In what follows, we will only
need to consider eqs.~(\ref{Einstein}, \ref{Maxwell}) in Minkowski
spacetime, although we will not necessarily work in Lorentzian
signature. Furthermore, we may also work in complexified Minkowski
space, whose line element is
\begin{equation}
ds^2=dt^2-dx^2-dy^2-dz^2,\quad t,x,y,z\in\mathbb{C}
\label{ds2C}
\end{equation}
n.b. the complexified coordinates, but not their complex conjugates,
appear.\\

An immediate use of the spinorial language is that it allows us to
classify different types of solution in electromagnetism and
gravity. Above, we noted that any symmetric spinor can be factorised
into 1-index principal spinors. These may be degenerate, and the
various different patterns of degeneracy allow for a classification of
different solutions. For the electromagnetic field strength spinor,
one has:
\begin{equation}
\phi_{AB}=\alpha_{(A}\beta_{B)},
\label{phiclassify}
\end{equation}
and there are then two different ``types'' of field strength spinor:
(i) those with distinct null directions ($\alpha_A\not\propto
\beta_A$); (ii) those with a degenerate null direction, so that
$\alpha_A\propto\beta_A$. The latter give rise to null electromagnetic
fields, given that the field strength spinor then satisfies
\begin{equation}
\phi_{AB}\phi^{AB}=\epsilon^{AC}\epsilon^{DB}\phi_{AB}\phi_{CD}
=\epsilon^{AC}\epsilon^{DB}\alpha_A\alpha_B\alpha_C\alpha_D=0.
\label{nullfield}
\end{equation}
For the Weyl tensor there are more possibilities. In general one has
\begin{equation}
\Psi_{ABCD}=\alpha_{(A}\beta_B\gamma_C\delta_{D)},
\label{Weylclassify}
\end{equation}
and one can classify the different types of Weyl spinor by their
pattern of degenerate null directions. If they are all different, this
is called a spinor of type $\{1,1,1,1\}$. Gradually making more of the
null directions degenerate leads to types $\{2,1,1\}$, $\{3,1\}$ and
$\{4\}$ (n.b the symmetrisation of the 1-index spinors in
eq.~(\ref{Weylclassify}) means that the {\it order} of the principal
spinors does not matter - only their degeneracy). A fifth possibility
is that there are two distinct pairs of null directions, denoted as
$\{2,2\}$. Finally, the Weyl spinor may be zero, which is written as
$\{-\}$. This reproduces the well-known {\it Petrov classification} of
GR solutions, which was first derived using tensorial methods. That
formalism used different labels to those used here, and we summarise
these in table~\ref{tab:Petrov}. We will use the Petrov labels from
now on. \\
\begin{table}
\begin{center}
\begin{tabular}{c|c}
Weyl type & Petrov label\\
\hline
$\{1,1,1,1\}$ & I\\
$\{2,1,1\}$ & II \\
$\{3,1\}$ & III\\
$\{4\}$ & N\\
$\{2,2\}$ & D\\
$\{-\}$ & O 
\end{tabular}
\caption{Different types of Weyl spinor classified by: (i) the
  pattern of degenerate principal null directions; (ii) the
  equivalent Petrov type.}
\label{tab:Petrov}
\end{center}
\end{table}

Having summarised spinorial notation and methods, we can now state the
Weyl double copy discussed in the introduction, which was first
presented in ref.~\cite{Luna:2018dpt}. Given an electromagnetic field
strength spinor $\phi_{AB}$, one may construct a Weyl spinor according
to the rule
\begin{equation}
\Psi_{ABCD}=\frac{1}{S}\phi_{(AB}\phi_{CD)},
\label{WeylDC}
\end{equation}
where $S$ is a scalar function. This procedure was argued to hold for
arbitrary type D vacuum spacetimes in ref.~\cite{Luna:2018dpt}, where
the scalar $S$ could then be found in particular examples by matching
both sides of eq.~(\ref{WeylDC}). All of these solutions have the
property that they linearise the Einstein equations. Thus, the
corresponding field strength and Weyl spinors may be taken to satisfy
eqs.~(\ref{Einstein}, \ref{Maxwell}) in Minkowski space i.e. such that
$\nabla^{AA'}$ corresponds to a flat-space derivative. The rule of
eq.~(\ref{WeylDC}) is given only for the anti-self-dual part of the
Weyl tensor. It is straightforward to write down the appropriate
generalisation for the self-dual part, and the above linearity
property then means that one may superpose these solutions to obtain
the complete Weyl tensor.\\

Equation~(\ref{WeylDC}) makes intuitive sense from our previous
discussion of principal null directions. Taking a field strength
spinor of the form of eq.~(\ref{phiclassify}), one obtains a Weyl
spinor of form
\begin{equation}
\Psi_{ABCD}\sim \alpha_{(A}\beta_B\alpha_C\beta_{D)},
\label{phi2}
\end{equation}
which is clearly of Petrov type D. This is not the only way to obtain
such a Weyl tensor. As already argued in ref.~\cite{Luna:2018dpt}, one
could also define a ``mixed'' Weyl double copy
\begin{equation}
\Psi_{ABCD}=\frac{1}{S}{\phi}_{(AB}{\tilde{\phi}}_{CD)},
\label{Weylmixed}
\end{equation}
involving two different electromagnetic spinors
\begin{equation}
\phi_{AB}=\alpha_A\alpha_B,\quad {\tilde{\phi}}_{CD}=\beta_C\beta_D.
\label{twospinors}
\end{equation}
This immediately suggests that one might be able to generalise the
Weyl double copy away from gravity solutions of type D: by taking
different patterns of principal null directions of both
electromagnetic spinors in eq.~(\ref{Weylmixed}), one may obtain a
variety of Petrov types, as summarised in table~\ref{tab:Petrov2}.
\begin{table}
\begin{center}
\begin{tabular}{c|c|c}
$\phi_{AB}$ & $\tilde{\phi}_{CD}$ & Petrov type\\
\hline
$\alpha_A\beta_B$ & $\gamma_C\delta_D$ & I\\
$\alpha_A\beta_B$ & $\alpha_C \gamma_D$ & II\\
$\alpha_A\alpha_B$ & $\beta_C \gamma_D$ & II\\
$\alpha_{A}\alpha_B$ & $\alpha_C\beta_D$ & III\\
$\alpha_{A}\alpha_B$ & $\alpha_C\alpha_D$ & N\\
$\alpha_{A}\alpha_B$ & $\beta_C\beta_D$ & D\\
$\alpha_{A}\beta_B$ & $\alpha_C\beta_D$ & D\\
\end{tabular}
\caption{Various combinations of principal null directions for two electromagnetic spinors lead to Weyl tensors of different Petrov types under the mixed Weyl double copy of eq.~(\ref{Weylmixed}).}
\label{tab:Petrov2}
\end{center}
\end{table}
We have yet to show whether such a double copy is meaningful, although
we will see later that this is indeed the case, provided one restricts
to linearised level in some cases.\\

Since its inception, the Weyl double copy has been used to study
certain topologically non-trivial electromagnetic solutions
(``Hopfions'')~\cite{Sabharwal:2019ngs}, and a tensorial formulation
of the relationship between the electromagnetic field strength and
Weyl tensor has been presented in
refs.~\cite{Alawadhi:2019urr,Alawadhi:2020jrv}, allowing a
generalisation to higher dimensions. An alternative approach to the
self-dual double copy which also has links to spinorial ideas may be
found in ref.~\cite{Elor:2020nqe}. Recently, a study of the Weyl
double copy properties of type N solutions has been presented in
ref.~\cite{Godazgar:2020zbv}.

\subsection{Twistors}
\label{sec:twistor}

In what follows, we will be concerned with deriving the Weyl double
copy using ideas from {\it twistor
  theory}~\cite{Penrose:1967wn,Penrose:1968me,Penrose:1972ia}. This is
a subject with an illustrious history spanning more than half a
century, and we cannot do justice to all of its many developments
here, but will concentrate on those aspects which are crucial for what
follows. Excellent reviews may be found in
refs.~\cite{Penrose:1986ca,Huggett:1986fs,Adamo:2017qyl}. We may start
by defining twistor space $\mathbb{T}$ as the set of solutions of the
{\it twistor equation}
\begin{equation}
\nabla_{A'}^{(A}\Omega^{B)}=0,
\label{twistoreq}
\end{equation}
whose general solution in Minkowksi space is
\begin{equation}
\Omega^A=\omega^A-ix^{AA'}\pi_{A'}.
\label{twistorsol}
\end{equation}
One may thus associate solutions of eq.~(\ref{twistoreq}) with
four-component objects (``twistors'') containing a pair of 2-spinors:
\begin{equation}
Z^\alpha=\left(\omega^A,\pi_{A'}\right).
\label{twistor}
\end{equation}
More explicitly, the index notation on the left is such that
\begin{equation}
(Z^0,Z^1,Z^2,Z^3)=\left(\omega^0,\omega^1,\pi_{0'},\pi_{1'}\right).
\label{twistor2}
\end{equation}
The ``location'' of a twistor in Minkowski space is defined to be the
region in which its associated spinor field $\Omega^A$ vanishes. From
eq.~(\ref{twistorsol}), this implies the {\it incidence relation}
\begin{equation}
\omega^A=ix^{AA'}\pi_{A'},
\label{incidence}
\end{equation}
where $x^{AA'}$ is the spinorial translation of a point in
space-time. Note that this condition is invariant under simultaneous
rescalings
\begin{equation}
\omega^A\rightarrow \lambda \omega^A,\quad \pi_{A'}\rightarrow \lambda
\pi_{A'},\quad \lambda \in\mathbb{C},
\label{rescale}
\end{equation}
so that twistors obeying the incidence relation are defined only up to
an arbitrary complex scale factor. They thus correspond to points in
{\it projective twistor space}, or $\mathbb{PT}$.  If we consider a
fixed point $x^{AA'}$ in eq.~(\ref{incidence}), this defines a line in
$\mathbb{PT}$. To see why, note that the constraint of
eq.~(\ref{incidence}) reduces the four complex parameters in a general
twistor to two, and a further complex parameter is removed by going to
the projective space. We therefore see that the relationship between
Minkowski space and $\mathbb{PT}$ is non-local~\footnote{Likewise,
  points in $\mathbb{PT}$ represent two (complex) parameter surfaces
  called {\it $\alpha$-planes} in complexified Minkowski space, or a
  null ray through a given point in real Minkowski space.}. A complex
line (with a point at infinity) can be mapped to a Riemann sphere.  \\

By considering the conjugate twistor equation
\begin{equation}
\nabla_A^{(A'}\Lambda^{B')}=0
\label{twistoreq2}
\end{equation}
whose general solution is 
\begin{equation}
\Lambda^{A'}=\mu^{A'}+ix^{AA'}\lambda_{A},
\label{twistorsol2}
\end{equation}
one may define {\it dual twistors}
\begin{equation}
W_\alpha=\left(\lambda_A,\mu^{A'}\right).
\label{dualtwistor}
\end{equation}
Projective dual twistor space is denoted by $\mathbb{PT}^*$. Similarly
to twistors, the location of a dual twistor in spacetime is defined by
the region where the associated spinor field $\Lambda^{A'}$ vanishes,
leading to an incidence relation
\begin{equation}
\mu^{A'}=-ix^{AA'}\lambda_A.
\label{incidence2}
\end{equation}
There is an inner product between twistors and dual twistors, defined
by
\begin{equation}
Z^\alpha W_\alpha=\omega^A\lambda_A+\mu^{A'}\pi_{A'},
\label{twistorprod}
\end{equation}
where we have defined components as in eqs.~(\ref{twistor},
\ref{dualtwistor}). These various objects have a number of very nice
mathematical properties. For example, one may show that the generators
of the conformal group act linearly on (dual) twistors, and that the
inner product of eq.~(\ref{twistorprod}) is conformally
invariant. This explains the ubiquity of twistor techniques in modern
research on scattering ampitudes, given that there are many situations
in which (dual-) (super-) conformal invariance can be made manifest,
particularly in more symmetric theories such as ${\cal N}=4$ SYM. \\

\begin{figure}
\begin{center}
\scalebox{0.6}{\includegraphics{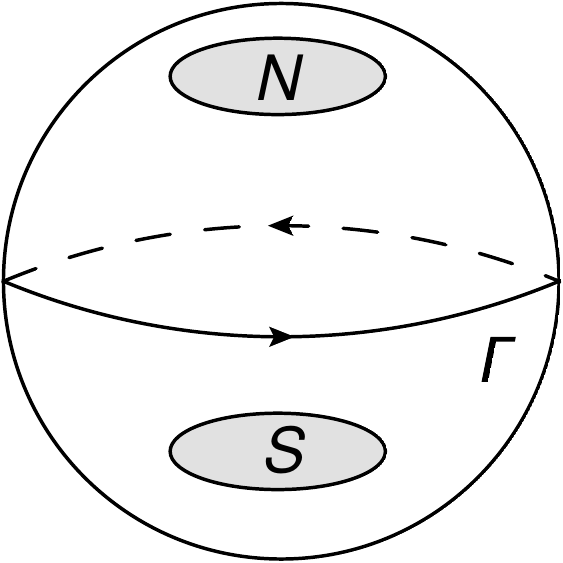}}
\caption{A contour $\Gamma$ divides the Riemann sphere into two
  hemispheres, with singularities in two regions $N$ and $S$.}
\label{fig:hemispheres}
\end{center}
\end{figure}

Another useful result from twistor theory is that there is a known
correspondence between solutions of the massless free field equations
of eq.~(\ref{massless}) in Minkowski spacetime, and certain contour
integrals in projective twistor space~\cite{Penrose:1968me}. This is
called the {\it Penrose transform}, and involves twistor
functions~\footnote{The functions $f(Z^\alpha)$ are actually
  representatives of cohomology classes, as we describe below.}
$f(Z^\alpha)$ (i.e. not involving the conjugates
$\bar{Z}^\alpha$). Given such a function with $Z^\alpha$ defined as in
eq.~(\ref{twistor}), we may define the integral
\begin{equation}
\phi_{A'B'\ldots C'}(x)=\frac{1}{2\pi i}\oint_{\Gamma}\pi_{E'}d\pi^{E'}
\pi_{A'}\pi_{B'} \ldots \pi_{C'}[\rho_x f(Z^\alpha)] ,
\label{Penrose}
\end{equation}
where the symbol $\rho_x$ denotes that we must restrict to the line in
$\mathbb{PT}$ corresponding to the spacetime point $x^{AA'}$. The
contour $\Gamma$ for this integral is defined on the related Riemann
sphere, and is well-defined only if the various poles appearing in
$f(Z^\alpha)$ can be separated from each other. We will see explicit
examples of functions and contours later on. Note that, for
eq.~(\ref{Penrose}) to make sense as an integral in projective twistor
space, the integrand (including the measure) must be homogeneous of
degree zero under rescalings $\pi_{A'}\rightarrow \lambda \pi_{A'}$
(or $Z^\alpha\rightarrow \lambda Z^\alpha$). This in turn implies that
the function $f(Z^\alpha)$ must have degree $(-n-2)$, where $n$ is the
number of indices appearing on the left-hand side. Clearly the field
on the left-hand side is symmetric in $\{A',B',\ldots, C'\}$, and thus
could potentially represent a solution of eq.~(\ref{massless}). The
proof that it does indeed do so is relatively straightforward. First,
one notes~\footnote{The fact that only the derivative with respect to
  $\omega^D$ arises in eq.~(\ref{masssol1}) and not its conjugate, is
  a consequence of $f(Z^\alpha)$ being a representative of a
  cohomology class. See ref.~\cite{Adamo:2017qyl} for a discussion of
  this issue from a different point of view.}
\begin{align}
\nabla_{DD'}[\rho_xf(Z^\alpha)]
&=\frac{\partial}{\partial x^{DD'}}f(ix^{AA'}\pi_{A'},\pi_{A'})\notag\\
&=i\pi_{D'}\rho_x\left[\frac{\partial f(Z^\alpha)}{\partial \omega^D}\right].
\label{masssol1}
\end{align}
Then from eq.~(\ref{Penrose}) we have
\begin{equation}
\nabla_{DD'}\phi_{A'B'\ldots C'}=
\frac{1}{2\pi}\oint_\Gamma \pi_{E'}d\pi^{E'}
\pi_{A'}\pi_{B'} \ldots \pi_{C'}\pi_{D'}
\rho_x\left[\frac{\partial f(Z^\alpha)}{\partial \omega^D}\right].
\label{masssol2}
\end{equation}
The right-hand side is symmetric in $A'$ and $D'$. Thus, contracting
both sides with $\epsilon^{A'D'}$ yields
\begin{equation}
\nabla^{A'}_D\phi_{A'B'\cdots C'}=0
\label{masssol3}
\end{equation}
as required. Note that there is considerable freedom on the right-hand
side of eq.~(\ref{Penrose}), due to being able to move the contour,
and also being able to modify the function $f(Z^\alpha)$ with
additional terms that vanish upon performing the contour
integration. To examine this in more detail, note that any contour
will separate the Riemann sphere into two parts, such that any
singularities are associated with two regions on either side of the
contour. Without loss of generality, we denote this as in
figure~\ref{fig:hemispheres} with the contour at the equator of the
sphere, and the singularities confined to regions $N$ and $S$ in the
northern and southern hemispheres respectively.
We must get the same answer for the integral of eq.~(\ref{Penrose}) if
we take the contour to enclose wither $N$ or $S$ (up to a change of
sign for differing orientation of the contour). However, we are free
to modify the twistor function according to the equivalence relation
\begin{equation}
f(Z^\alpha)\rightarrow f(Z^\alpha)+f_N(Z^\alpha)+f_S(Z^\alpha),
\label{fshift}
\end{equation}
where $f_N(Z^\alpha)$ ($f_S(Z^\alpha)$) contains singularities only in
the northern (southern) hemisphere. In carrying out the contour
integral for either of $f_{N,S}(Z^\alpha)$, one may simply choose to
enclose the hemisphere that is free of singularities, thus obtaining
zero. \\

The above remarks make clear that the Penrose transform map from
twistor space to spacetime is many-to-one. In more mathematical terms,
this may be described in terms of {\it sheaf cohomology groups} (see
e.g. refs.~\cite{Penrose:1986ca,Huggett:1986fs} for a review), where
the notation ${\rm H}^1(\mathbb{PT},{\cal O}(m))$ represents, roughly
speaking, the equivalence class of holomorphic functions of
homogeneity $m$ in projective twistor space (avoiding the regions $N$
and $S$), modulo the redefinitions of eq.~(\ref{fshift}). The above
Penrose transform then amounts to an isomorphism between massless
fields of helicity $n$ in spacetime, and the cohomology
group~\footnote{In practice, one takes this isomorphism on a suitable
  open subset of $\mathbb{PT}$ e.g. that corresponding to positive or
  negative frequency fields in spacetime.} ${\rm
  H}^1(\mathbb{PT},{\cal O}(-n-2))$. \\

Above, we have given the Penrose transform for primed fields. A
similar formula may be written for unprimed fields, if one instead
uses holomorphic functions defined on projective dual twistor space:
\begin{equation}
\psi_{AB\cdots D}(x)=\frac{1}{2\pi i}\oint_\Gamma \lambda_E \,d\lambda^E\,
\lambda_A\,\lambda_B\ldots\lambda_D\left[\rho_x
f(W_\alpha)\right].
\label{Penrosedual}
\end{equation}
Analogously to eq.~(\ref{Penrose}), this constitutes an isomorphism
between massless fields in spacetime with helicity $-n$, and the
cohomology group ${\rm H}^1(\mathbb{PT}^*,{\cal O}(-n-2))$. In early
works on twistor theory, it was seen as unsatisfactory that both
twistors and dual twistors were needed to define fields of arbitrary
helicity. This in turn led to the alternative procedure for negative
helicity fields~\cite{Hughston:1979pg}
\begin{equation}
\psi_{AB\ldots C}(x)=\frac{1}{2\pi i}\oint \pi_{E'}d\pi^{E'}
\rho_x\left[
\frac{\partial}{\partial \omega^A}
\frac{\partial}{\partial \omega^B}\ldots
\frac{\partial}{\partial \omega^C}
f(Z^\alpha)\right],
\label{Penrose2}
\end{equation}
in which the spinors $\omega^A$ are defined in eq.~(\ref{twistor}). In
this formula, the twistor function must now have homogeneity $(n-2)$
for consistency. In terms of the sheaf cohomology groups mentioned
above, one has an isomorphism between helicity $-n$ massless fields in
spacetime and elements of the cohomology group ${\rm
  H}^1(\mathbb{PT},{\cal O}(n-2))$. From now on, we will loosely use
the term {\it function} to imply a given representative of the
appropriate cohomology class, returning to a discussion of the latter
in section~\ref{sec:objection}.\\

Each of eqs.~(\ref{Penrose}, \ref{Penrosedual}, \ref{Penrose2}) is
defined for Minkowski space only. In gauge theory or gravity, they
thus give rise to field configurations that satisfy the linearised
Yang-Mills or Einstein equations. This does not bother us for the Weyl
double copy, all previous examples of which have been linear, but
exact, solutions. However, we must bear this in mind when considering
generalisations. Full non-linear generalisations of the Penrose
transform exist for the (anti-)self-dual sectors of Yang-Mills
theory~\cite{Ward:1977ta} and gravity~\cite{Penrose:1976jq}. Also,
even the linear Penrose transform above is more general than it
seems. Conformal invariance is manifest in twistor space, so that we
expect that the fields obtained from eqs.~(\ref{Penrose})
and~(\ref{Penrose2}) can also be transformed to an arbitrary
conformally flat spacetime. In fact, the massless free field equation
is known to be conformally invariant~\cite{Penrose:1987uia}, so this
is indeed the case. There is a subtlety, however, regarding scalar
fields. From eq.~(\ref{Penrose}), we expect these to be given by
contour integrals of the form
\begin{equation}
\phi(x)=\frac{1}{2\pi i}\oint_{\Gamma}\pi_{E'}d\pi^{E'}
[\rho_x f(Z^\alpha)].
\label{Penrose3}
\end{equation}
We can longer form an equation like eq.~(\ref{massless}), as there is
nothing for either of the spinor indices in the covariant derivative
$\nabla^{AA'}$ to contract with. However, one can instead show that
eq.~(\ref{Penrose}) is (in conformally flat spacetimes) a solution of
the conformally invariant wave equation
\begin{equation}
\left(\Box+\frac{R}{6}\right)\phi=0,
\label{wave}
\end{equation}
where $R$ is the Ricci scalar. \\

The above-mentioned many-to-one property of the Penrose transform of
eq.~(\ref{Penrose}) means that it is not possible to write a unique
inverse transform that fixes a twistor function $f(Z^\alpha)$ from a
given spacetime field. However, there are some tricks for formulating
representative twistor functions for spacetime fields possessing
certain properties. A particularly useful one is the observation made
in ref.~\cite{Penrose:1986ca}, relating the poles of twistor functions
to principal spinors in spacetime. First, note that the factorisation
property of symmetric spinors means that if a given $n$-index spinor
has a $k$-fold principal spinor $\xi_{A'}$, it will vanish if
contracted with $(n-k+1)$ factors of $\xi_{A'}$, but not if only
$(n-k)$ factors are contracted. To see this, we may write such a
spinor's decomposition into principal spinors as
\begin{equation}
\phi_{A'B'\dots F'}=\underbrace{\xi_{(A'}\xi_{B'}\ldots \xi_{C'}}_{k\ \rm{factors}}
\underbrace{\alpha_{D'}\beta_{E'}
\ldots \gamma_{F')}}_{(n-k)\ \rm{factors}},
\label{contract1}
\end{equation}
so that contracting this with $(n-k)$ factors of $\xi^{A'}$ gives 
\begin{equation}
\xi^{D'}\xi^{E'}\ldots \xi^{F'}\phi_{A'B'\ldots F'}
=[\alpha \xi][\beta\xi]\ldots [\gamma\xi]\xi_{(A'}\xi_{B'}\ldots
\xi_{C')},\quad [\lambda\mu]\equiv \lambda_{A'}\mu^{A'}.
\label{contract2}
\end{equation}
The various prefactors on the right-hand side are manifestly non-zero,
given that none of the spinors
$\{\alpha_A',\beta_{B'}\ldots\gamma_{C'}\}$ can be proportional to
$\xi_{A'}$, if they are distinct principal spinors. Contracting with a
further factor of $\xi^{A'}$ immediately gives a factor of
$[\xi\xi]=0$, thus proving the assertion made above. Now consider the
spacetime field defined by eq.~(\ref{Penrose}), in the case where the
twistor function $f(Z^\alpha)$ has a pole of order $m$ (where $m\leq
n$)\footnote{If the twistor function has a pole of higher order than
  $n$, enclosed by $\Gamma$, then the associated spinor is not even a
  principal spinor.}, which occurs when $\pi_{A'}\propto
\eta_{A'}$. Contracting eq.~(\ref{Penrose}) with $m$ factors of
$\eta^{A'}$ gives
\begin{equation}
\underbrace{\eta^{A'}\eta^{B'}\ldots \eta^{C'}}_{m\ {\rm factors}}
\underbrace{\phi_{A'B'\ldots C'D'\ldots F'}}_{n\ {\rm indices}}
(x)=\frac{1}{2\pi i}\oint_{\Gamma}\pi_{E'}d\pi^{E'}
[\pi\eta]^m \pi_{C'}\ldots \pi_{F'}[\rho_x f(Z^\alpha)].
\label{polestructure1}
\end{equation}
The factor $[\pi\eta]^m$ in the integrand will kill the $m^{\rm
  th}$-order pole as $\pi_{A'}\rightarrow \eta_{A'}$, such that the
contour integral is zero. By the above remarks, and given that this
will not occur if one contracts with only $(m-1)$ factors of
$\eta^{A'}$, we thus find that in the Penrose transform of
eq.~(\ref{Penrose}), the field $\phi_{A'B'\ldots F'}$ has at least a
$(n-m+1)$-fold principal spinor $\eta_{A'}$, if the twistor function
$f(Z^\alpha)$ has a single $m^{\rm th}$-order pole as
$\pi_{A'}\rightarrow\eta_{A'}$, enclosed by $\Gamma$. Furthermore, if
this pole remains present for varying $x^{AA'}$, then the twistor
function must have the general form~\cite{Penrose:1986ca}
\begin{equation}
f(Z^\alpha)=\theta_m(Z^\alpha)\{\chi(Z^\alpha)\}^{-m},
\label{Zform}
\end{equation}
where $\theta_m(Z^\alpha)$, $\chi(Z^\alpha)$ are homogeneous and
holomorphic, $\theta_m(Z^\alpha)$ is regular at the $m$-fold pole we
are discussing, and $\chi(Z^\alpha)$ has a simple zero. \\

\section{A twistorial derivation of the Weyl double copy}
\label{sec:Weyl}

In the previous section, we have reviewed various aspects of spinors
and twistors, culminating in the Penrose transform of
eq.~(\ref{Penrose}), and its related results of
eqs.~(\ref{Penrosedual}, \ref{Penrose2}). In this section, we show how
these results can be used to derive the Weyl double copy of
ref.~\cite{Luna:2018dpt}, presented here in eq. (\ref{WeylDC}). We will
focus on the conjugate form of eq.~(\ref{WeylDC}), involving fields
with primed indices. From the remarks of section~\ref{sec:twistor},
this means that we will be concerned with maps from $\mathbb{PT}$ to
spacetime, rather than $\mathbb{PT^*}$. Changing notation for later
convenience, the general (mixed) Weyl double copy may be written as
\begin{equation}
\phi_{A'B'C'D'}=\frac{1}{\phi}\phi^{(1)}_{(A'B'}\phi^{(2)}_{C'D')}.
\label{WeylDC2}
\end{equation}
Here $\phi$ satisfies the scalar wave equation, and
$\phi^{(1,2)}_{A'B'}$ are two electromagnetic spinors. Then
$\phi_{A'B'C'D'}$ satisfies the spin-2 case of the massless free field
equation of eq.~(\ref{massless}), and thus represents a self-dual Weyl
spinor. As discussed in the introduction, the Weyl double copy is
related to previous exact classical double copies that have appeared
in the literature where overlap exists e.g. the Kerr-Schild double
copy of ref.~\cite{Monteiro:2014cda}. It has been argued to hold for
general type D vacuum spacetimes in ref.~\cite{Luna:2018dpt}, and also
for type N vacuum spacetimes in ref.~\cite{Godazgar:2020zbv}. However,
a number of questions remain:
\begin{enumerate}
\item The form of eq.~(\ref{WeylDC2}) involves a certain product of
  spinorial quantities, with symmetrisation over the indices. Is there
  a deeper explanation of where this form comes from?
\item Similarly, why should there be an exact double copy relating
  fields in position space? The original BCJ double copy for
  amplitudes~\cite{Bern:2010ue,Bern:2010yg} involves products in {\it
    momentum space}, which would be expected to lead to a convolution
  in position space. Indeed, there are alternative double copy
  formalisms that have exactly this
  property~\cite{Anastasiou:2014qba,Anastasiou:2015vba,Anastasiou:2016csv,Anastasiou:2017nsz,Cardoso:2016ngt,Anastasiou:2017taf,LopesCardoso:2018xes,Anastasiou:2018rdx,Borsten:2020bgv,Cardoso:2016amd,Borsten:2020xbt,Borsten:2020zgj}.
\item Does the double copy apply to Petrov types other than type D or
  N? As already discussed in section~\ref{sec:spinorial}, the form of
  eqs.~(\ref{WeylDC}, \ref{WeylDC2}) would seem to apply for arbitrary
  Petrov types, although it is not of course guaranteed that the
  resulting spinorial quantity on the left-hand side will satisfy the
  spin-2 massless free field equation.
\item Can the Weyl double copy be extended to curved spacetime
  backgrounds?  Preliminary work for the Kerr-Schild double copy
  implies that exact results are possible in conformally flat
  spacetimes~\cite{Bahjat-Abbas:2017htu,Carrillo-Gonzalez:2017iyj}.
\end{enumerate}
Reference~\cite{White:2020sfn} recently presented (in a very brief
form) a twistor space procedure for deriving the type D Weyl double
copy, and also gave an example of a more general solution (of Petrov
type III) that could be obtained from eq.~(\ref{WeylDC2}). We discuss
this, with full details that are missing in ref.~\cite{White:2020sfn},
in the following sections.

\subsection{Twistor space picture}
\label{sec:twistorspace}

Consider a two holomorphic twistor functions $f_{\rm
  EM}^{(1,2)}(Z^\alpha)$ of homogeneity $-4$, and a further holomorphic
twistor function $f(Z^\alpha)$ of homogeneity $-2$. By the Penrose
transform described here in section~\ref{sec:twistor}, these will
necessarily respectively correspond to electromagnetic spinors
$\phi^{(1,2)}_{A'B'}$ and a scalar field $\phi$ in spacetime. One may
then form a product
\begin{equation}
f_{\rm grav.}(Z^\alpha)=
\frac{f^{(1)}_{\rm EM}(Z^\alpha)\,f^{(2)}_{\rm EM}(Z^\alpha)}{f(Z^\alpha)},
\label{WeylDC3}
\end{equation}
such that the function on the left-hand side necessarily has
homogeneity $-6$, and thus potentially corresponds to a spacetime
field solving the spin-2 massless free field equation i.e. to a
self-dual (linearised) gravity solution~\footnote{The reader may
  object at this point that we are multiplying together twistor
  functions. We return to this in section~\ref{sec:objection}.}. All
we have done here is use known properties of the Penrose
transform. However, the above remarks imply that there might be some
sort of relationship between gauge, gravity and scalar fields in
spacetime that corresponds to the product of eq.~(\ref{WeylDC3}). By
choosing suitable functions on the right-hand side of
eq.~(\ref{WeylDC3}), we can generate a particular spacetime
relationship. We now show that for a suitable choice of twistor
functions, this spacetime relationship is precisely the type D Weyl
double copy of eq.~(\ref{WeylDC2}).\\

To find the appropriate functions, we may rely on the result of
eq.~(\ref{Zform}), namely that a spacetime field possessing a
$(n-m+1)$-fold principal spinor must have a $m$-fold pole in twistor
space. A type D Weyl spinor has a pair of 2-fold degenerate principal
spinors, such that the twistor function $f(Z^\alpha)$ must have two
triple poles. This in turn implies that the function $\chi(Z^\alpha)$
in eq.~(\ref{Zform}) has two simple zeros, and thus corresponds to
some quadratic form~\footnote{It is important here that the Weyl
  spinor we are seeking has no more than two distinct principal
  spinors: more than two, and the contour $\Gamma$ in the Penrose
  transform would be enclosing more than one pole in at least one
  hemisphere of the Riemann sphere, thus invalidating the remarks
  leading to eq.~(\ref{Zform}). We discuss this point in more detail
  in section~\ref{sec:Petrov}.}
\begin{equation}
\chi(Z^\alpha)=Q_{\alpha\beta}Z^\alpha Z^\beta,
\label{chiform}
\end{equation}
for some constant $Q_{\alpha\beta}$. Let us also define $\theta_m$ in
eq.~(\ref{Zform}) to be the simple combinatorial factor
\begin{equation}
\theta_m=\frac{1}{m!},
\label{thetam}
\end{equation}
for reasons that will become clear. Then, putting everything together,
we may define a family of twistor functions
\begin{equation}
f_m(Z^\alpha)=\frac{1}{m!}\left[Q_{\alpha\beta}Z^\alpha Z^\beta
\right]^{-m},
\label{fmdef}
\end{equation}
and our claim is that this will produce a type D Weyl tensor (for
$m=3$), that is related to an electromagnetic field strength ($m=2$)
and scalar field ($m=1$). To show that this is indeed true, we may
carry out the Penrose transform in each case by choosing homogeneous
coordinates
\begin{equation}
\pi_{A'}=(1,\xi),\quad \xi\in\mathbb{C}
\label{coords}
\end{equation}
(n.b. we are allowed to do this, given that the integrand of the
Penrose transform has homogeneity zero under rescalings of
$\pi_{A'}$). The quadratic form in eq.~(\ref{chiform}) is to be
evaluated when the incidence relation of eq.~(\ref{incidence}) is
obeyed, namely for
$Z^\alpha=(ix^{AA'}\pi_{A'},\pi_{A'})$. Substituting
eq.~(\ref{coords}) then implies
\begin{equation}
\chi(Z^\alpha)=N^{-1}(x)(\xi-\xi_1)(\xi-\xi_2)
\label{chiform2}
\end{equation}
in general, where $\xi_i\equiv \xi_i(x)$ expresses the location of a
pole in terms of the variable $\xi$, and $N^{-1}(x)$ is an overall
normalisation factor. The dependence of the incidence relation on
$x^{AA'}$ gives rise to the spacetime dependence of $\{N,\xi_i\}$. The
measure of the Penrose transform becomes simply 
\begin{equation}
\pi^{E'}d\pi_{E'}=d\xi,
\label{measure}
\end{equation}
and we can now consider each value of $m$ in turn. For the scalar case
of $m=1$, the Penrose transform reduces to
\begin{align}
\phi&=\frac{1}{2\pi i}\oint_{\Gamma}d\xi \frac{N(x)}{(\xi-\xi_1)(\xi-\xi_2)}
\notag\\
&=\frac{N(x)}{2\pi i}\oint_{\Gamma}d\xi \frac{1}{(\xi-\xi_1)(\xi-\xi_2)}.
\label{m=1}
\end{align}
The contour $\Gamma$ is defined on the Riemann sphere of $\xi$, such
that it separates the poles at $\xi=\xi_1$ and $\xi=\xi_2$. We may
thus evaluate the contour integral by enclosing only one of these
poles, which we take to be $\xi=\xi_1$. Cauchy's theorem then tells us
that
\begin{equation}
\oint_\Gamma \frac{d\xi}{(\xi-\xi_1)(\xi-\xi_2)}=\frac{2\pi i}{\xi_1-\xi_2},
\label{Cauchy1}
\end{equation}
so that we have 
\begin{align}
\phi(x)&=\frac{N(x)}{\xi_1-\xi_2}.
\label{m=1b}
\end{align}
Next, we have the case $m=2$, which gives the Penrose transform
\begin{align}
\phi_{A'B'}=\frac{N^2(x)}{2\pi i}\frac{1}{2!}\oint_\Gamma d\xi 
\frac{(1,\xi)_{A'}(1,\xi)_{B'}}{(\xi-\xi_1)^2(\xi-\xi_2)^2}.
\label{m=2}
\end{align}
We choose to evaluate the integral for each choice of $\{A',B'\}$ by
again enclosing the pole at $\xi=\xi_1$. Recalling that if a function
$f(z)$ has an $n^{\rm th}$-order pole at $z=c$, the residue associated
with the latter is
\begin{equation}
{\rm res}(f,c)=\frac{1}{(n-1)!}\lim_{z\rightarrow c}\frac{d^{n-1}}{dz^{n-1}}
\left[(z-c)^n f(z)\right],
\label{residue}
\end{equation}
one may then verify the following:
\begin{equation}
\frac{1}{2\pi i}\oint_\Gamma d\xi \frac{\xi^n}{(\xi-\xi_1)^2(\xi-\xi_2)^2}=
-\frac{1}{(\xi_1-\xi_2)^3}\times
\begin{cases}
2, & n=0,\\
(\xi_1+\xi_2), & n=1,\\
2\xi_1\xi_2, & n=2,
\end{cases}
\end{equation}
which in turn implies 
\begin{equation}
\phi_{0'0'}=-\frac{N^2(x)}{(\xi_1-\xi_2)^3},\quad
\phi_{0'1'}=\phi_{1'0'}=-\frac{N^2(x)}{(\xi_1-\xi_2)^3}\frac{1}{2}
(\xi_1+\xi_2),\quad
\phi_{1'1'}=-\frac{N^2(x)}{(\xi_1-\xi_2)^3}\xi_1\xi_2.
\label{phiABsol}
\end{equation}
Alternatively, one may express $\phi_{A'B'}$ directly in terms of its
principal spinors. Defining
\begin{equation}
\alpha_{A'}=(1,\xi_1),\quad \beta_{A'}=(1,\xi_2),
\label{abdef}
\end{equation}
eq.~(\ref{phiABsol}) is equivalent to
\begin{equation}
\phi_{A'B'}=-\frac{N^2(x)}{(\xi_1-\xi_2)^3}\alpha_{(A'}\beta_{B')}.
\label{phiABsol2}
\end{equation}
Finally, we can examine the case $m=3$, which produces a Weyl spinor
as follows:
\begin{equation}
\phi_{A'B'C'D'}=\frac{N^3(x)}{2\pi i}\frac{1}{3!}\oint_\Gamma d\xi
\frac{(1,\xi)_{A'}(1,\xi)_{B'}(1,\xi)_{C'}(1,\xi)_{D'}}
{(\xi-\xi_1)^3(\xi-\xi_2)^3}.
\label{weylsol1}
\end{equation}
Again using eq.~(\ref{residue}), one finds
\begin{align}
\frac{1}{2\pi i}\oint_{\Gamma}d\xi \frac{\xi^n}
{(\xi-\xi_1)^3(\xi-\xi_2)^3}
=\frac{1}{(\xi_1-\xi_2)^5}\times
\begin{cases}
6, & n=0,\\
3(\xi_1+\xi_2), & n=1,\\
\xi_1^2+4\xi_1\xi_2+\xi_2^2, & n=2,\\
3\xi_1\xi_2(\xi_1+\xi_2), & n=3,\\
6\xi_1^2\xi_2^2, & n=4.
\end{cases}
\label{weylsol2}
\end{align}
Substituting this into eq.~(\ref{weylsol1}), one finds that the result
may be written as
\begin{align}
\phi_{A'B'C'D'}&=\frac{N^3(x)}{(\xi_1-\xi_2)^5}\alpha_{(A'}\beta_{B'}
\alpha_{C'}\beta_{D')}\notag\\
&=\left[\frac{N(x)}{(\xi_1-\xi_2)}\right]^{-1}
\phi_{(A'B'}\phi_{C'D')},
\label{weylsol3}
\end{align}
where in the second line we have recognised the spin-1 massless field
(i.e. an electromagnetic field strength tensor) of
eq.~(\ref{phiABsol2}), and used the symmetrisation property
\begin{equation}
\Psi_{((A'B')(C'D'))}=\Psi_{(A'B'C'D')},
\label{symsym}
\end{equation}
which holds for an arbitrary spinor $\Psi_{A'B'C'D'}$. We may
recognise the prefactor in eq.~(\ref{weylsol3}) as the inverse of the
scalar field of eq.~(\ref{m=1b}). Thus, our choice of twistor
functions has provided a scalar, electromagnetic field strength and
(linearised) Weyl tensor satisfying
\begin{equation}
\phi_{A'B'C'D'}=\frac{1}{\phi}\phi_{(A'B'}\phi_{C'D')}.
\label{WeylDCtwistor}
\end{equation}
This is precisely the (primed version of) the Weyl double copy of
eq.~(\ref{WeylDC}). Some further comments are in order:
\begin{itemize}
\item Our choice of the combinatorial factor in eq.~(\ref{thetam}) is
  to reproduce the normalisation of the Weyl double copy as given in
  ref.~\cite{Luna:2018dpt}, and one needs such a factor whenever
  higher-order poles are present. However, the scalar function $S(x)$
  is itself only defined up to a constant factor in that paper, so
  that this is not strictly necessary. Furthermore, we are working
  with linearised field equations, so that any constant factor is
  possible.
\item The twistor story explains why there is a classical double copy
  in position space, given that the Penrose transform links twistor
  functions with spacetime fields. Furthermore, the somewhat
  mysterious form of eq.~(\ref{WeylDC2}) is now also explained. 
\item In the original formulation of the Weyl double copy, there was
  no clear prescription for fixing the scalar function $S$. Here, we
  see that it naturally arises as the scalar field $\phi$ obtained
  from the $m=1$ case of eq.~(\ref{fmdef}).
\item Equation~(\ref{WeylDCtwistor}) is limited to linearised
  equations of motion only. However, this is not a problem for the
  spacetimes considered in ref.~\cite{Luna:2018dpt}, all of which
  linearise the Einstein equations, so that the linearised double copy
  can be promoted to an exact statement. 
\item The twistor formalism is in principal conformally invariant (at
  least in those cases in which the twistor functions do not involve
  the {\it infinity twistors} for a given spacetime, that break
  conformal invariance). This means that the Weyl double copy should
  immediately extend to conformally flat spacetimes, thus formalising
  a preliminary observation made in
  ref.~\cite{Bahjat-Abbas:2017htu}. One could, for example, take a
  given set of scalar, electromagnetic and gravity fields that are
  linked by the Weyl double copy, and conformally transform them
  directly to a desired spacetime, thus achieving a double copy on a
  curved background. We leave a full investigation of this interesting
  possibility to future work.
\end{itemize}
In order for the above to constitute a full derivation of the Weyl
double copy, it must be the case that all possible vacuum type D
spacetimes can be obtained using twistor functions of the form of
eq.~(\ref{fmdef}). That this is indeed the case has been argued by
Haslehurst and Penrose in ref.~\cite{Haslehurst}, and general
arguments may also be given. Type D vacuum solutions are distinguished
by the presence of two distinct shear-free null geodesic
congruences. All such congruences (in Minkowski space) can be obtained
as the zero sets of twistor functions, by a result known as the {\it
  Kerr theorem} (see also
e.g. refs.~\cite{Penrose:1986ca,Huggett:1986fs}). In the present case,
these twistor functions are precisely those appearing in the
denominator of eq.~(\ref{fmdef}).

\subsection{Example: Schwarzschild \& Taub-NUT}
\label{sec:Schwarzschild}

A canonical example is that of the Schwarzschild black hole, which is
not (anti-)self-dual by itself. However, it is
known~\cite{Alawadhi:2019urr,Huang:2019cja,Banerjee:2019saj} that
duality transformations map out the parameter space of a general
Taub-NUT solution with Schwarzschild mass $M$ and NUT charge
$N$. Thus, if we restrict to the self-dual part of the Weyl tensor
only, we will obtain self-dual Taub-NUT with a fixed relationship
between $M$ and $N$. To obtain this using the above construction, one
may take a family of functions as in eq.~(\ref{fmdef}), where a
suitable choice for $Q_{\alpha\beta}$ is~\footnote{We use the
  self-dual analogue of the anti-self-dual twistor function presented
  in refs.~\cite{Quadrille,Sparling,Hughston:1979tq}.}
\begin{equation}
Q_{\alpha\beta}=\frac{1}{2}\left(\begin{array}{rrrr}0&0&0&-1\\
0&0&1&0\\
0&1&0&0\\
-1&0&0&0\end{array}\right),
\label{SchwarzschildQ}
\end{equation}
so that the quadratic form in eq.~(\ref{chiform}) becomes
\begin{align}
\chi&=Z^1Z^2-Z^0Z^3\notag\\
&=\omega^1\pi_{0'}-\omega^0\pi_{1'}.
\label{chiSZ}
\end{align}
This must be evaluated subject to the incidence relation of
eq.~(\ref{incidence}), where
\begin{equation}
x^{AA'}=\frac{1}{\sqrt{2}}\left(\begin{array}{cc}
t+z & x+iy\\ x-iy & t-z\end{array}\right).
\label{xaa'}
\end{equation}
With $\pi_{A'}$ given by eq.~(\ref{coords}), we then find
\begin{equation}
\chi=-\frac{i}{\sqrt{2}}\left(\xi^2(x+iy)+2z\xi-(x-iy)\right).
\label{chiSZ2}
\end{equation}
This has roots at 
\begin{equation}
\xi_{1,2}=\frac{-z\pm r}{(x+iy)}=\frac{x-iy}{z\mp r},
\quad r=\sqrt{x^2+y^2+z^2},
\label{xi12SZ}
\end{equation}
and comparing eq.~(\ref{chiSZ2}) with eq.~(\ref{chiform2}), we find
\begin{equation}
N(x)=\frac{i\sqrt{2}}{(x+iy)}.
\label{NSZ}
\end{equation}
From eq.~(\ref{m=1b}), the biadjoint scalar function $\phi$ associated
with this solution is given by
\begin{equation}
\phi=\frac{i}{r\sqrt{2}}.
\label{phiSZ}
\end{equation}
This agrees with the function $S(x)$ presented in
ref.~\cite{Luna:2018dpt}~\footnote{Reference~\cite{Luna:2018dpt}
  actually presents results for the Kerr solution, but this reduces to
  the Schwarzschid solution when the angular momentum parameter $a$ is
  taken to zero.}, up to an overall normalisation constant. However,
their function $S(x)$ is itself only defined up to an overall
constant, so this is not a problem. \\

The field strength and Weyl spinors associated with the
electromagnetic and gravity solutions will have the form of
eqs.~(\ref{phiABsol2}, \ref{weylsol3}), where the two principal
spinors in the present case are
\begin{equation}
\alpha_{A'}=\left(1,\frac{x-iy}{z+r}\right),\quad
\beta_{A'}=\left(1,\frac{x-iy}{z-r}\right).
\label{alphabetaSZ}
\end{equation}
We may relate these to the Kerr-Schild double copy of
ref.~\cite{Monteiro:2014cda} as follows. In a real spacetime, we
construct the principal null directions corresponding to the real
spinors by combining each of the latter with their complex conjugates
and the appropriate Infeld-van-der-Waerden symbols:
\begin{equation}
k^{(1)}_\mu=\alpha_{A'}\bar{\alpha}_A\sigma_\mu^{AA'},\quad
k^{(2)}_\mu=\beta_{A'}\bar{\beta}_A\sigma_\mu^{AA'}.
\label{k12def}
\end{equation}
Using the spinors of eq.~(\ref{abdef}), one finds
\begin{equation}
k^{(i)}_\mu=\frac{1}{\sqrt{2}}\left(
1+|\xi_i|^2,\xi_i+\xi_i^*,i(\xi_i-\xi_i^*),1-|\xi_i|^2\right),
\label{kisol1}
\end{equation}
such that the explicit forms of eq.~(\ref{alphabetaSZ}) yield
\begin{equation}
k^{(1,2)}_\mu\propto \frac{r\sqrt{2}}{r\pm z}
\left(1,\pm\frac{x}{r},\pm\frac{y}{r},\pm\frac{z}{r}\right).
\label{k12sol2}
\end{equation}
The proportional sign here arises from the fact that the spinors of
eq.~(\ref{abdef}) were defined in projective space, and thus can be
renormalised by a position-dependent factor. This allows us to remove
the prefactor on the right-hand side of eq.~(\ref{k12sol2}), and one
recovers the two possible choices of Kerr-Schild vector $k_\mu$ for
the Schwarzschild spacetime~\cite{Monteiro:2014cda}. \\

\subsection{Examples of general Petrov type}
\label{sec:Petrov}

Above, we have seen that it is possible to derive the type D Weyl
double copy, by choosing appropriate functions in the twistor space
product of eq.~(\ref{WeylDC3}). However, the spacetime form of the
(mixed) double copy, eq.~(\ref{Weylmixed}), is not intrinsically
limited to producing type D solutions only. Thus the question
naturally arises as to whether arbitrary Petrov types are
possible. This was briefly considered in ref.~\cite{White:2020sfn},
which presented examples with Petrov types N and III (see also
ref.~\cite{Godazgar:2020zbv} for a discussion of Petrov type N). These
examples utilised a particularly well-studied class of holomorphic
twistor functions, namely {\it elementary states} (see
e.g. ref.~\cite{Penrose:1986ca}), which consist of ratios of factors
of the form $(A_\alpha Z^\alpha)$, where $A_\alpha$ is a constant dual
twistor. Elementary states were originally intended as alternatives to
plane wave states, for the purposes of examining scattering processes
in twistor space. However, they were recently reconsidered from a
different point of view, as the twistor functions associated with
certain topologically non-trivial electromagnetic
fields. Reference~\cite{Dalhuisen:2012zz} pointed out that the field
associated with the zeroes of the twistor function $(A_\alpha
Z^\alpha)$ is an electromagnetic {\it Hopf knot}, any pair of whose
electric (or magnetic) field lines are
linked. Reference~\cite{Swearngin:2013sks} generalised this further,
by considering the Penrose transform of the family of twistor
functions
\begin{equation}
f_{h}(Z^\alpha)=(A_\alpha Z^\alpha)^{-1}(B_\beta Z^\beta)^{-2h-1},
\label{fhpdef}
\end{equation}
where $h$ is the helicity of the resulting field in spacetime.  We may
write the dual twistors in eq.~(\ref{fhpdef}) as~\footnote{In a slight
  abuse of notation we have used $A$ and $B$ to denote dual twistors,
  as well as their associated Weyl-spinors. However, the nature of the
  index in each case makes this notation unambiguous.}
\begin{equation}
A_\alpha=(A_A,A^{A'}),\quad B_\alpha=(B_A,B^{A'}).
\label{ABdef}
\end{equation}
Furthermore, we follow ref.~\cite{Swearngin:2013sks} in defining the
calligraphic quantities
\begin{equation}
 \rho_x[A_\alpha Z^\alpha]\equiv{\cal A}^{A'}\pi_{A'},\quad
\rho_x[B_\beta Z^\beta]\equiv {\cal B}^{B'}\pi_{B'},
\label{ABcaldef}
\end{equation}
such that
\begin{equation}
{\cal A}^{A'}=ix^{AA'}A_A+A^{A'},
\label{ABcaldef2}
\end{equation}
and similarly for ${\cal B}^{B'}$. Then the corresponding solutions of
the massless free field equations were found to be
\begin{equation}
\phi_{A'_1\ldots A'_{2h}}(x)=\left(\frac{2}{\Omega|x-y|^2}\right)^{2h+1}
{\cal A}_{A'_1}\ldots {\cal A}_{A'_{2h}},
\label{Hopfions}
\end{equation}
where
\begin{equation}
\Omega=A_B B^B,\quad y^{AA'}=i\frac{B^A A^{A'}-A^A B^{A'}}
{A_B B^B}.
\label{Omegadef}
\end{equation}
The spin-1 and spin-2 versions of eq.~(\ref{Hopfions}) correspond to a
null electromagnetic spinor and type N Weyl spinor respectively. The
general field is referred to as a {\it spin-$N$ Hopfion}, given that
its spacetime topology is related to the well-known {\it Hopf
  fibration}. \\

A further generalisation was presented in
ref.~\cite{Thompson:2014pta}, which considered the family of twistor
functions
\begin{equation}
f(Z^\alpha)=\frac{(C_\gamma Z^\gamma)^{h(n_p-1)}(D_\delta Z^\delta)^{h(n_t-1)}}
{(A_\alpha Z^\alpha)(B_\beta Z^\beta)^{h(n_p+n_t)+1}}.
\label{Hopfgen}
\end{equation}
Here $h$ is the helicity as before, and
$n_p,n_t\in\mathbb{Z}$. Defining calligraphic spinorial quantities as
in eq.~(\ref{ABcaldef}), the corresponding spacetime fields via the
Penrose transform were found to be
\begin{equation}
\phi_{A'_1\ldots A'_{2h}}(x)=\frac{({\cal A}_{C'}{\cal C}^{C'})^{h(n_p-1)}
({\cal A}_{D'}{\cal D}^{D'})^{h(n_t-1)}}
{({\cal A}_{B'}{\cal B}^{B'})^{h(n_p+n_t)+1}}{\cal A}_{(A'_1}\ldots
{\cal A}_{A'_{2h})}.
\label{torusknots}
\end{equation}
These fields again have a non-trivial topology: in the electromagnetic
case, electric field lines correspond to {\it torus knots}, where
$n_t$ and $n_p$ denote the toroidal and poloidal winding numbers. A
similar geometry (involving gravitoelectric field lines, defined from
the parity-even part of the Weyl tensor) can be ascertained in the
spin-2 case. Again, the electromagnetic field strength is null, and
the Weyl spinor is type N. Reference~\cite{Thompson:2014owa} sought to
generalise this, by constructing gravitational solutions with
different Petrov types. In particular, the following Penrose transform
was noted~\footnote{The results of ref.~\cite{Thompson:2014owa} appear
  not to include an overall combinatorial factor, which we have
  explicitly instated here.}:
\begin{align}
\frac{1}{(A_\alpha Z^\alpha)^{1+a}(B_\alpha Z^\alpha)^{1+b}}
&\rightarrow \frac{{\cal N}_{ab}}{[{\cal A}{\cal B}]^{a+b+1}}
{\cal A}_{(A'_1}\ldots{\cal A}_{A'_b}{\cal B}_{A'_{b+1}}\ldots
{\cal B}_{A'_{2h})}\notag\\
&={\cal N}_{ab}\left(\frac{2}{\Omega|x-y|^2}\right)^{a+b+1}
{\cal A}_{(A'_1}\ldots{\cal A}_{A'_b}{\cal B}_{A'_{b+1}}\ldots
{\cal B}_{A'_{2h})},
\label{knots1}
\end{align}
where
\begin{equation}
{\cal N}_{ab}=(-1)^a\left(\begin{array}{c}a+b\\a\end{array}\right).
\label{Nabdef}
\end{equation}
The Weyl double copy properties of torus knots were considered using
methods similar to refs.~\cite{Tod,Berman:2018hwd} in
ref.~\cite{Sabharwal:2019ngs}, although the form of the biadjoint
field was not explicitly discussed there. However, the above results
fit very nicely into the twistor space picture for obtaining the Weyl
double copy. Starting with eqs.~(\ref{fhpdef}, \ref{Hopfions}), one
may take the cases with $h=0$, $h=1$ and $h=2$ as the scalar,
electromagnetic and gravity functions in the twistor space product of
eq.~(\ref{WeylDC3}), such that the corresponding spacetime fields are
\begin{equation}
\phi=\frac{2}{\Omega|x-y|^2},\quad
\phi_{A'B'}=\left(\frac{2}{\Omega|x-y|^2}\right)^3 {\cal A}_{(A'}
{\cal A}_{B')},\quad
\phi_{A'B'C'D'}=\left(\frac{2}{\Omega|x-y|^2}\right)^5 {\cal A}_{(A'}
{\cal A}_{B'}{\cal A}_{C'}{\cal A}_{D')},
\label{Hopfion}
\end{equation}
from which it is straightforward to verify
eq.~(\ref{Weylmixed}). A similar analysis can be carried out for
eqs.~(\ref{Hopfgen}) and~(\ref{torusknots}). \\

Next, consider eq.~(\ref{knots1}). If $a=b=0$, one obtains the scalar
field of eq.~(\ref{Hopfion}). One may construct twistor functions of
homogeneity $-4$ by choosing $(a,b)=(1,1)$ or $(0,2)$, leading to the two
respective electromagnetic spinors
\begin{equation}
\phi^{(1,1)}_{A'B'}=-2\left(\frac{2}{\Omega|x-y|^2}\right)^3{\cal A}_{(A'}
{\cal B}_{B')},\quad
\phi^{(0,2)}_{A'B'}=\left(\frac{2}{\Omega|x-y|^2}\right)^3{\cal A}_{(A'}
{\cal A}_{B')}.
\label{emspinors}
\end{equation}
Using these in the mixed Weyl double copy of eq.~(\ref{Weylmixed}),
one can generate a number of different Weyl spinors:
\begin{align}
\phi^{(1,1)\times(1,1)}_{A'B'C'D'}&=4\left(\frac{2}
{\Omega|x-y|^2}\right)^5{\cal A}_{(A'}{\cal A}_{B'}{\cal B}_{C'}
{\cal B}_{D')},\notag\\
\phi^{(1,1)\times (0,2)}_{A'B'C'D'}&=-2\left(\frac{2}
{\Omega|x-y|^2}\right)^5{\cal A}_{(A'}{\cal A}_{B'}{\cal A}_{C'}
{\cal B}_{D')},\\
\phi^{(0,2)\times (0,2)}_{A'B'C'D'}&=\left(\frac{2}
{\Omega|x-y|^2}\right)^5{\cal A}_{(A'}{\cal A}_{B'}{\cal A}_{C'}
{\cal A}_{D')},
\label{weylchoices}
\end{align}
which is entirely consistent with the rule for combining the
corresponding functions in twistor space. The first and third of these
examples are Petrov type D and N respectively. However, the second (as
already noted in ref.~\cite{Thompson:2014owa}) is Petrov type III,
thus going beyond the original formulation of the Weyl double copy in
ref.~\cite{Luna:2018dpt}.\\

We may go further than the above (and the results of
ref.~\cite{White:2020sfn}) by seeking Weyl double copy examples with
Petrov types I and II, where again we may rely on elementary
states. However, we will see that this leads to a generalisation of
the Weyl double copy formula of eq.~(\ref{WeylDC2}). Let us illustrate
this with the simpler case of type II solutions. Consider the twistor
function
\begin{equation}
f_{\rm grav.}^{\rm (II)}=\frac{C_\gamma Z^\gamma}{(A_\alpha Z^\alpha)^3
(B_\beta Z^\beta)^4}\equiv\frac{\mathcal{C}^{E'}\pi_{E'}}{(\mathcal{A}^{G'}\pi_{G'})^{3}(\mathcal{B}^{H'}\pi_{H'})^{4}},
\label{WeylType2}
\end{equation}
where $A_\alpha$, $B_\alpha$ and $C_\alpha$ are constant dual
twistors, and we have defined calligraphic spinors as in
eq.~(\ref{ABcaldef}). This has homogeneity $-6$ and thus represents a
gravity solution in spacetime, where the appropriate Penrose transform
can be written as
\begin{equation}
\Psi^{\rm (II)}_{A'B'C'D'}=\mathcal{C}^{E'}\phi_{A'B'C'D'E'}^{(2,3)},
\label{PsiIIdef}
\end{equation}
where
\begin{align}
\phi_{A'B'C'D'E'}^{(2,3)}&=
\frac{1}{2\pi i}\oint_{\Gamma}\frac{\pi_{A'}\pi_{B'}\pi_{C'}\pi_{D'}\pi_{E'}}
{(\mathcal{A}^{G'}\pi_{G'})^{3}(\mathcal{B}^{H'}\pi_{H'})^{4}}\pi_{I'}
d\pi^{I'},\notag\\
&=\frac{10}{[\mathcal{AB}]^{6}}\mathcal{A}_{(A'}\mathcal{A}_{B'}
\mathcal{A}_{C'}\mathcal{B}_{D'}\mathcal{B}_{E')},
\label{phiABCDE}
\end{align}
and we have used eq.~(\ref{knots1}) in the second line. To see why
eq.~(\ref{PsiIIdef}) yields a type II Weyl spinor, note that one may
rewrite the spinorial quantity on the right-hand side of
eq.~(\ref{phiABCDE}) as
\begin{equation}
\mathcal{A}_{(A'}\mathcal{A}_{B'}\mathcal{A}_{C'}\mathcal{B}_{D'}\mathcal{B}_{E')}=\frac{1}{5}\left[3\mathcal{A}_{E'}\mathcal{B}_{(A'}+2\mathcal{B}_{E'}\mathcal{A}_{(A'}\right]\mathcal{A}_{B'}\mathcal{A}_{C'}\mathcal{B}_{D')},
\label{sym2}
\end{equation}
such that the Weyl spinor of eq.~(\ref{PsiIIdef}) may be rewritten as
\begin{equation}
\Psi_{A'B'C'D'}^{\rm (II)}=\frac{2}{[\mathcal{AB}]^{5}}\mathcal{A}_{(A'}
\mathcal{A}_{B'}\mathcal{B}_{C'}\mathcal{F}_{D')},
\label{Type3410}
\end{equation}
where
\begin{equation}
{\cal F}_{A'}=3\frac{[\mathcal{CA}]}{[\mathcal{AB}]}
\mathcal{B}_{A'}+2\frac{[\mathcal{CB}]}{[\mathcal{AB}]}\mathcal{A}_{A'}.
\label{Fcaldef}
\end{equation}
Equation~(\ref{Type3410}) is then manifestly of type II as required,
and to show that it may be obtained by a Weyl double copy, we must
find twistor functions that can be substituted in eq.~(\ref{WeylDC3})
so as to reproduce eq.~(\ref{WeylType2}). In fact, we need only use
the twistor functions related to the electromagnetic Hopfions discussed
above. To this end, consider the homogeneity $-4$ functions related to
the two electromagnetic spinors (\ref{emspinors}). These are
\begin{align}
f^{(0,2)}_{\rm EM}&=\frac{1}{(A_\alpha Z^\alpha)(B_\beta Z^\beta)^3}=
\frac{1}{[\pi {\cal A}][\pi{\cal B}]^3},\notag\\
f^{(1,1)}_{\rm EM}&=\frac{1}{(A_\alpha Z^\alpha)^2(B_\beta Z^\beta)^2}=
\frac{1}{[\pi {\cal A}]^2[\pi {\cal B}]^2},
\label{f12EMdef}
\end{align}
where we have used the notations of eqs.~(\ref{contract2},
\ref{ABcaldef}), as well as the homogeneity $-2$ function
\begin{equation}
f^{(0,0)}=\frac{1}{(A_\alpha Z^\alpha)(B_\beta Z^\beta)}=\frac{1}
{[\pi{\cal A}][\pi {\cal B}]}. 
\label{ftypeII}
\end{equation}
Upon applying the Schouten identity
\begin{equation}
[{\cal C}{\cal A}][\pi{\cal B}]+[{\cal B}{\cal C}][\pi{\cal A}]
+[{\cal A}{\cal B}][\pi{\cal C}]=0,
\label{Schouten}
\end{equation}
one may then rewrite eq.~(\ref{WeylType2}) as~\footnote{This factorization is not unique. The other possibility is $f_{\rm grav.}^{\rm (II)}=\frac{1}{f^{(0,0)}}f_{\rm EM}^{\rm (0,2)}\left(
\frac{[{\cal C}{\cal A}]}{[{\cal A}{\cal B}]}f_{\rm EM}^{\rm (2,0)}
-\frac{[{\cal C}{\cal B}]}{[{\cal A}{\cal B}]}f_{\rm EM}^{\rm (1,1)}
\right)$, where $f_{\rm EM}^{\rm (2,0)}$ is given in eq. (\ref{f3def}).}
\begin{equation}
f_{\rm grav.}^{\rm (II)}=\frac{1}{f^{(0,0)}}f_{\rm EM}^{\rm (1,1)}\left(
-\frac{[{\cal C}{\cal B}]}{[{\cal A}{\cal B}]}f_{\rm EM}^{\rm (0,2)}
+\frac{[{\cal C}{\cal A}]}{[{\cal A}{\cal B}]}f_{\rm EM}^{\rm (1,1)}
\right).
\label{WeylType2b}
\end{equation}
The spacetime field corresponding to eq.~(\ref{WeylType2b}) is a Weyl
tensor of the form

\begin{equation}
\Psi_{A'B'C'D'}^{\rm (II)}=\frac{1}{\phi}\left[3
\frac{[{\cal C}{\cal A}]}{[{\cal A}{\cal B}]} \phi^{(0,2)}_{(A'B'}
\phi^{(1,1)}_{C'D')}-4\frac{[{\cal C}{\cal B}]}{[{\cal A}{\cal B}]}
\phi^{(1,1)}_{(A'B'}\phi^{(1,1)}_{C'D')}\right],
\label{WeylDCsum}
\end{equation}
in agreement with eqs.~(\ref{Type3410}, \ref{Fcaldef}). Here, we have
combined the electromagnetic functions to make a gravity function in
twistor space, and only then carried out the Penrose transform. This
correctly keeps track of combinatorial factors resulting from the
multiplicities of the poles in twistor space, and the result of our
procedure is that we obtain a generalised double copy formula
(eq.~(\ref{WeylDCsum})), containing a sum of two distinct terms, each
with the structure of eq.~(\ref{WeylDC2})~\footnote{We have chosen to
  keep out factors of spinor brackets in eq.~(\ref{WeylDCsum}), but
  could just as easily have absorbed these into the definitions of the
  electromagnetic spinors on the right-hand side.}. \\

We may carry out a similar analysis for Petrov type I, by considering
e.g. the twistor function
\begin{equation}
f_{\rm grav.}^{\rm (I)}=\frac{(\mathcal{C}^{E'}\pi_{E'})^{2}}{(\mathcal{A}^{G'}\pi_{G'})^{4}(\mathcal{B}^{H'}\pi_{H'})^{4}},
\label{WeylType1}
\end{equation}
whose Penrose transform yields 
\begin{eqnarray}
\Psi_{A'B'C'D'}^{\rm (I)}
&=&\mathcal{C}^{E'}\mathcal{C}^{F'}\phi_{A'B'C'D'E'F'}^{(3,3)},
\label{Type4420}
\end{eqnarray}
with
\begin{eqnarray}
\phi_{A'B'C'D'E'F'}^{(3,3)}&=&\frac{1}{2\pi i}\oint\frac{\pi_{A'}\pi_{B'}\pi_{C'}\pi_{D'}\pi_{E'}\pi_{F'}}{(\mathcal{A}^{G'}\pi_{G'})^{4}(\mathcal{B}^{H'}\pi_{H'})^{4}}\pi_{I'}d\pi^{I'}\nonumber\\
&=&-\frac{20}{[\mathcal{AB}]^{7}}\mathcal{A}_{(A'}\mathcal{A}_{B'}\mathcal{A}_{C'}\mathcal{B}_{D'}\mathcal{B}_{E'}\mathcal{B}_{F')}.
\label{phiABCDEF}
\end{eqnarray}
But
\begin{equation}
\mathcal{A}_{(A'}\mathcal{A}_{B'}\mathcal{A}_{C'}\mathcal{B}_{D'}\mathcal{B}_{E'}\mathcal{B}_{F')}=\frac{1}{5}\left[3\mathcal{A}_{(E'}\mathcal{B}_{F')}\mathcal{A}_{(A'}\mathcal{B}_{B'}+\mathcal{A}_{E'}\mathcal{A}_{F'}\mathcal{B}_{(A'}\mathcal{B}_{B'}+\mathcal{B}_{E'}\mathcal{B}_{F'}\mathcal{A}_{(A'}\mathcal{A}_{B'}\right]\mathcal{A}_{C'}\mathcal{B}_{D')},
\end{equation}
such that eq.~(\ref{Type4420}) gives
\begin{align}
\Psi_{A'B'C'D'}^{\rm (I)}&=-\frac{4}{[\mathcal{AB}]^{7}}\left[3[\mathcal{CA}][\mathcal{CB}]\mathcal{A}_{(A'}\mathcal{B}_{B'}+[\mathcal{CA}]^{2}\mathcal{B}_{(A'}\mathcal{B}_{B'}+[\mathcal{CB}]^{2}\mathcal{A}_{(A'}\mathcal{A}_{B'}\right]\mathcal{A}_{C'}\mathcal{B}_{D')}\notag\\
&=-\frac{1}{[{\cal A}{\cal B}]^5}{\cal A}_{(A'}{\cal B}_{B'}{\cal D}_{C'}{\cal E}_{D')},
\label{Type4420b}
\end{align}
where
\begin{align}
{\cal D}_{C'}=(3-\sqrt{5})\frac{[{\cal A}{\cal C}]}{[{\cal A}{\cal B}]}
{\cal B}_{C'}+\frac{2[{\cal B}{\cal C}]}{[{\cal A}{\cal B}]}{\cal A}_{C'},
\notag\\
{\cal E}_{D'}=(3+\sqrt{5})\frac{[{\cal A}{\cal C}]}{[{\cal A}{\cal B}]}
{\cal B}_{D'}+\frac{2[{\cal B}{\cal C}]}{[{\cal A}{\cal B}]}{\cal A}_{C'}.
\label{DEcaldef}
\end{align}
Equation~(\ref{Type4420b}) is manifestly of type I as required. As for
the type II example of eq.~(\ref{WeylDCsum}), it may be written as a
superposition of pure Weyl double copies. By repeated application of
eq.~(\ref{Schouten}), we may rewrite the twistor function of
eq.~(\ref{WeylType1}) as~\footnote{Similar to the type II case, this factorization is not unique. The other possibility is $
f_{\rm grav.}^{\rm (I)}=\frac{1}{f^{(0,0)}}f^{(1,1)}_{EM}
\left(\frac{[\mathcal{CA}]^{2}}{[\mathcal{AB}]^{2}}f^{(2,0)}_{EM}+\frac{[\mathcal{CB}]^{2}}{[\mathcal{AB}]^{2}}f^{(0,2)}_{EM}-2\frac{[\mathcal{CA}][\mathcal{CB}]}{[\mathcal{AB}]^{2}}f^{(1,1)}_{EM}\right)$.}
\begin{eqnarray}
f_{\rm grav.}^{\rm (I)}=\frac{1}{f^{(0,0)}}\left(\frac{[\mathcal{CB}]}{[\mathcal{AB}]}f^{(0,2)}_{EM}-\frac{[\mathcal{CA}]}{[\mathcal{AB}]}f^{(1,1)}_{EM}\right)
\left(-\frac{[\mathcal{CA}]}{[\mathcal{AB}]}f^{(2,0)}_{EM}+\frac{[\mathcal{CB}]}{[\mathcal{AB}]}f^{(1,1)}_{EM}\right),
\label{Weyltype1b}
\end{eqnarray}
where we have introduced a third homogeneity $-4$ function
\begin{equation}
f_{\rm EM}^{\rm (2,0)}=\frac{1}{(A_\alpha Z^\alpha)^3(B_\beta Z^\beta)}=
\frac{1}{[\pi{\cal A}]^3[\pi{\cal B}]},
\label{f3def}
\end{equation}
with the respective electromagnetic spinor
\begin{equation}
\phi^{(2,0)}_{A'B'}=-\left(\frac{2}{\Omega|x-y|^2}\right)^3{\cal B}_{(A'}
{\cal B}_{B')}.
\end{equation}
Expanding and transforming each term separately to position space, one
finds 
\begin{align}
\Psi^{\rm (I)}_{A'B'C'D'}&=\frac{1}{\phi}
\left[\frac{3}{2}
\frac{[{\cal C}{\cal A}][{\cal C}{\cal B}]}
{[{\cal A}{\cal B}]^2}\left(\phi^{\rm (0,2)}_{(A'B'}
\phi^{\rm (2,0)}_{C'D')}-
\phi^{\rm (1,1)}_{(A'B'}\phi^{\rm (1,1)}_{C'D')}
\right)\right.\notag\\
&\left.\quad\quad+\frac{[{\cal C}{\cal B}]^2}
{[{\cal A}{\cal B}]^2}\phi^{\rm (0,2)}_{(A'B'}
\phi^{\rm (1,1)}_{C'D')}
-\frac12\frac{[{\cal C}{\cal A}]^2}
{[{\cal A}{\cal B}]^2}\phi^{\rm (1,1)}_{(A'B'}
\phi^{\rm (2,0)}_{C'D')}\right].
\label{Phi1expand}
\end{align}
As in the previous type II example, this is a sum of pure double copy
terms. However, in both cases, there is considerable choice in how one
presents the final results. Returning to the simpler type II example
of eq.~(\ref{WeylDCsum}), one may define the alternative
electromagnetic spinor
\begin{equation}
\Phi_{A'B'}=3\frac{[{\cal C}{\cal A}]}
{[{\cal A}{\cal B}]}\phi^{(0,2)}_{A'B'}
-4\frac{[{\cal C}{\cal B}]}{[{\cal A}{\cal B}]}\phi^{(1,1)}_{A'B'},
\label{Phi4def}
\end{equation}
which is guaranteed to solve the massless free field equation given
that the two terms on the right-hand side are themselves solutions,
and thus may be linearly superposed. Equation~(\ref{WeylDCsum}) then
becomes
\begin{equation}
\Psi_{A'B'C'D'}^{\rm (II)}=\frac{1}{\phi}\Phi_{(A'B'}
\phi^{\rm (1,1)}_{C'D')},
\label{WeylPhi4}
\end{equation}
which is of pure Weyl double copy form. The reader may be worried that
there are apparently different double copy formulae that can be
written down that relate different electromagnetic solutions to a
given gravity solution. However, this is in fact neither surprising
nor profound. The Penrose transform used here is limited to the
linearised gauge and gravity theories only, as are our examples of
gauge / gravity solutions, such that the ambiguity in associating a
given gravity solution with a given pair of electromagnetic solutions
is precisely that associated with being able to linearly superpose the
latter. Notably, the individual terms in eqs.~(\ref{WeylDCsum},
\ref{Phi1expand}) correspond to Weyl spinors of restricted Petrov
type, such that the superpositions involved correspond to the known
property that, in the linearised theory, one may superpose solutions
to create different Petrov types. We have thus succeeded in providing
Weyl double copy examples of more general Petrov type, but in a rather
artificial way. One may therefore question the utility of the twistor
approach (and indeed the Weyl double copy in general) for these
solutions. However, what the twistor framework does is provide an
interesting way to classify possible double copy formulae, in that the
problem of finding the different single copies of a given Weyl tensor
amounts to obtaining the different factorizations of the related
twistorial function. It also provides a motivation for why particular
solutions may be interesting even at linearised level (e.g. the
identification of elementary states with Hopfions and torus
knots~\cite{Dalhuisen:2012zz,Swearngin:2013sks,Thompson:2014owa,Thompson:2014pta}). It
would of course be very interesting to find examples of arbitrary
Petrov type where {\it exact} -- or at the very least non-linear --
solutions are related.

\subsection{A possible objection}
\label{sec:objection}

In the previous sections, we have outlined a derivation of the Weyl
double copy, that relies on a certain product of holomorphic twistor
functions in projective twistor space. However, this should rightly
incur the wrath of any sensible twistor theorist: as we discussed in
section~\ref{sec:twistor}, the ``functions'' we have discussed above
are not actually functions, but representatives of cohomology
classes. Each spin-$n$ (positive helicity) massless free field in
spacetime corresponds to a particular element (cohomology class) from
the group ${\rm H}^1(\mathbb{PT},-n-2)$, and the interpretation of the
product of eq.~(\ref{WeylDC3}) is then not at all clear~\footnote{We
  thank Prof. Edward Witten for comments leading to the present
  discussion.}.\\

In more pedestrian terms, the twistor function corresponding to a
given spacetime field is not unique, but may be redefined by adding
functions whose singularities lie on only one side of the contour
$\Gamma$ on the Riemann sphere corresponding to a given spacetime
point. Then, the product of eq.~(\ref{WeylDC3}) that is needed to
obtain the Weyl double copy in position space appears incompatible
with the ability to perform equivalence relations according to
eq.~(\ref{fshift}), in that the order of these operations does not
commute. To illustrate this point, it is sufficient to consider
redefining the twistor functions in the numerator of
eq.~(\ref{WeylDC3}), according to
\begin{equation}
f_{\rm EM}^{(i)}(Z^\alpha)\rightarrow\tilde{f}_{\rm EM}^{(i)}(Z^\alpha)
\equiv f_{\rm EM}^{(i)}(Z^\alpha)+\chi(Z^\alpha),
\label{fshift2}
\end{equation}
where $\chi(Z^\alpha)$ has homogeneity $-4$, and contains poles either
in the northern or southern hemisphere when restricted to the
Riemann sphere of spacetime point $x$, but not both. By
construction, the functions $\tilde{f}^{(i)}(Z^\alpha)$ give rise to
the same electromagnetic spinors $\phi^{(i)}_{A'B'}(x)$ as the
functions $f^{(i)}_{\rm EM}(Z^\alpha)$. However, forming the product
of eq.~(\ref{WeylDC3}) for the redefined functions leads to the
twistor function
\begin{equation}
\frac{\tilde{f}^{(1)}_{\rm EM}(Z^\alpha)
\tilde{f}^{(2)}_{\rm EM}(Z^\alpha)}{f(Z^\alpha)}=
\frac{f^{(1)}_{\rm EM}(Z^\alpha)
f^{(2)}_{\rm EM}(Z^\alpha)}{f(Z^\alpha)}+
\sum_{i=1}^2\frac{\chi(Z^\alpha)f_{\rm EM}^{(i)}}{f(Z^\alpha)}
+\frac{\chi^2(Z^\alpha)}{f(Z^\alpha)}.
\label{ftildeprod}
\end{equation}
Both the second and third terms on the right-hand side have
homogeneity $-6$, and thus the right-hand side gives rise to a solution
of the massless spin 2 free field equation in spacetime. However, the
second term on the right-hand side involves the original functions
$f_{\rm EM}^{(i)}(Z^\alpha)$, and thus will have poles in both the
northern and southern hemispheres of the Riemann sphere of
$x$. Recognising the first term on the right-hand side as our original
gravity function in twistor space, we thus see that
eq.~(\ref{ftildeprod}) does not correspond to an equivalence relation
of the form of eq.~(\ref{fshift}). Consequently, the transformation on
the right-hand side will gives rise to a different spacetime gravity
solution in general.\\

If we instead take given representative members of the equivalence
class of functions for $(f_{\rm EM}^{(i)}, f(Z^\alpha))$ and form the
product of eq.~(\ref{WeylDC3}), we are indeed free to make
redefinitions according to eq.~(\ref{fshift}). That is, the
transformations
\begin{equation}
\frac{f^{(1)}_{\rm EM}(Z^\alpha)
f^{(2)}_{\rm EM}(Z^\alpha)}{f(Z^\alpha)}\rightarrow
\frac{f^{(1)}_{\rm EM}(Z^\alpha)
f^{(2)}_{\rm EM}(Z^\alpha)}{f(Z^\alpha)}+
f_N(Z^\alpha)+f_S(Z^\alpha)
\label{fprodshift}
\end{equation}
do indeed yield equivalent gravity solutions. However, we are then
faced with the puzzle of how to pick out what these representative
members are meant to be, given that all possible choices of the
classes of function entering eq.~(\ref{WeylDC3}) are meant to be
equivalent!\\

The above puzzle, whilst interesting, does not appear to pose an
obstacle to deriving the Weyl double copy in spacetime. All one has to
do to achieve the latter is to pick suitable representatives from each
cohomology class, chosen by construction so as to obtain the type D
Weyl double copy of eq.~(\ref{WeylDC2}). Put another way, one only
needs to verify the following statement: {\it for particular elements
  (cohomology classes) from the groups ${\rm H}^1(\mathbb{PT},-2)$,
  ${\rm H}^1(\mathbb{PT},-4)$ and ${\rm H}^1(\mathbb{PT},-6)$, a
  representative of each class exists such that the corresponding
  spacetime fields obey eq.~(\ref{WeylDC2})}. This is a much weaker
statement than requiring a complete map between the classes themselves
i.e. intepreting the product of eq.~(\ref{WeylDC3}) as providing a
general map:
\begin{equation}
{\rm H}^1(\mathbb{PT},-2)\times {\rm H}^1(\mathbb{PT},-4)
\times{\rm H}^1(\mathbb{PT},-4)\rightarrow {\rm H}^1(\mathbb{PT},-6),
\label{comap}
\end{equation}
which may or may not be achievable. The validity of the weaker
statement above is demonstrated explicitly in
section~\ref{sec:twistorspace}, but whether or not anything more
general can be said is certainly worth investigating, as it is clearly
related to central questions regarding the validity and scope of the
double copy, including to exact solutions of arbitrary Petrov type.\\

We note also that from a physics point of view, the situation is
highly reminiscent of the well-known BCJ double copy for (quantum)
scattering amplitudes~\cite{Bern:2010ue,Bern:2010yg}, in which gravity
amplitudes are expressed as a sum of terms, each involving a product
of kinematic factors $\{n_i\}$ obtained from gauge theory
amplitudes. These numerators are gauge-dependent, but such that the
total amplitude is gauge-invariant. The double copy structure is not
manifest in arbitrary gauges, and one must make {\it generalised gauge
  transformations} (including also field definitions in general) in
order to put the numerators into a specific ``BCJ-dual'' form, so that
the double copy can be carried out. This problem already occurs at
tree-level, and if a given set of such numerators is subjected to a
gauge transformation
\begin{equation}
n_i\rightarrow n_i+\delta_i
\label{nishift}
\end{equation}
for some $\delta_i$, the double copy formula will generate unwanted
terms in the gravity amplitude, that threaten the gauge-invariance of
the latter. It is possible to set up the double copy in a more
gauge-invariant manner, but at the expense of having to introduce
additional correction terms on the gravity side, to cancel out the
unwanted contributions~\cite{Bern:2017yxu}. Although the situation
here is not exactly identical (i.e. the equivalence transformations of
eq.~(\ref{fshift}) do not correspond to spacetime gauge
transformations), it may well be that some similar procedure in
twistor space can be defined, so that full invariance with respect to
equivalence transformations is made manifest. Any such procedure
presumably faces the additional barrier of having to be interpretable
in sheaf cohomological terms, but there is again hope. For example,
products of twistor space cohomology classes have been discussed in
earlier literature regarding twistor diagrams for scattering
amplitudes (see e.g.~\cite{Hughston:1979pg,Bailey:1990qn} for
reviews). Some of these techniques may be adaptable to the present
case of classical solutions, and there may also be existing results
from the algebraic geometry literature regarding maps similar to those
required here (although we do not know of anything at the time of
writing).\\

Throughout, we have been discussing twistor cohomology classes using
the language of sheaf cohomology (or alternatively {\it \u{C}ech
  cohomology}, which is an appropriate approximation). However,
another formulation of the Penrose transform exists, in which the
twistor functions become differential forms, and are to be interpreted
as {\it Dolbeault cohomology classes} (see
e.g.~\cite{Woodhouse:1985id} for a review). That this is equivalent to
the above approach follows from known isomorphisms between \u{C}ech
and Dolbeault cohomology groups. It would certainly be interesting to
try to reformulate our derivation of the Weyl double copy in the
Dolbeault approach, as this is clearly related to whether the double
copy has a genuinely twistorial interpretation.

\subsection{The Weyl double copy for anti-self-dual fields}
\label{sec:anti-self-dual}

In the previous sections, as in ref.~\cite{White:2020sfn}, we have
addressed the Weyl double copy for self-dual fields i.e. those with
primed spinor indices. In this section, we extend this discussion to
anti-self-dual fields. As reviewed in section~\ref{sec:twistor}, there
are two Penrose transforms one may consider for anti-self-dual
fields. The first (eq.~(\ref{Penrosedual})) simply consists of
replacing twistors with dual twistors, and it is straightforward to
see that the derivation of the type D Weyl double copy in terms of
anti-self-dual fields proceeds similarly to the case of self-dual
fields discussed above. That is, one may consider the family of
functions
\begin{equation}
\tilde{f}_m=\frac{1}{m!}\left[Q^{\alpha\beta} W_\alpha W_\beta\right]^{-m},
\label{fbarmdef}
\end{equation}
with $Q^{\alpha\beta}$ a constant matrix. In the Penrose transform,
this is to be evaluated subject to the incidence relation of
eq.~(\ref{incidence2}), and one may choose homogeneous coordinates
\begin{equation}
\lambda_A=(1,\eta)
\label{lambdacoords}
\end{equation}
such that the quadratic form appearing on the right-hand side of
eq.~(\ref{fbarmdef}) may be written as
\begin{equation}
\tilde{\chi}\equiv \rho_x\left[Q^{\alpha\beta} W_\alpha W_\beta\right]
=\tilde{{\cal N}}^{-1}(x)(\eta-\eta_1(x))(\eta-\eta_2(x)),
\label{barchidef}
\end{equation}
for some spacetime-dependent functions $\tilde{N}$ and
$\eta_i$. Carrying out the Penrose transforms for $m\in\{1,2,3\}$
yields spacetime fields
\begin{align}
S=\frac{\tilde{N}(x)}{\eta_1-\eta_2},\quad
\phi_{AB}=-\frac{\tilde{N}^2(x)}{(\eta_1-\eta_2)^3}\alpha_{(A}\beta_{B)},
\quad
\Psi_{ABCD}=\frac{\tilde{N}^3(x)}{(\eta_1-\eta_2)^5}\alpha_{(A}
\beta_{B}\alpha_{C}\beta_{D)},
\label{dualfields}
\end{align}
obeying the Weyl double copy formula of eq.~(\ref{WeylDC}). \\

One may also consider using the (non-dual) twistor space Penrose
transform of eq.~(\ref{Penrose2}), but the complication then arises of
how to form a product in twistor space (i.e. before or after the
derivatives are applied). In section~\ref{sec:twistorspace}, each
quantity entering the twistor space product must be interpretable by
itself as corresponding to a spacetime field, after restriction to a
given spacetime point.  In eq.~(\ref{Penrose2}), the restriction to a
given spacetime point happens {\it after} the function $f(Z^\alpha)$
has already been differentiated, which suggests that we define a
twistor-space product in terms of differentiated quantities:
\begin{equation}
f_{AB\ldots C}=\frac{\partial}{\partial\omega^A}
\frac{\partial}{\partial\omega^B}\ldots
\frac{\partial}{\partial\omega^C}
f(Z^\alpha).
\label{findexdef}
\end{equation}
A twistorial double copy for anti-self-dual fields can then be written
as
\begin{equation}
f^{\rm grav.}_{ABCD}=\frac{f^{\rm EM}_{(AB}f^{\rm EM}_{CD)}}{f}.
\label{ASD_DC}
\end{equation}
We can at least show that such a relationship holds in particular
cases. For example, a suitable function to be entered into
eq.~(\ref{Penrose2}) for the (anti-self-dual) Coulomb solution
is~\cite{Quadrille}
\begin{equation}
f^{\rm EM}=\log\left(\frac{Q}{P}\right),
\label{FEMcoulomb}
\end{equation}
with
\begin{equation}
\label{P_and_Q}
\begin{aligned}
P=&Z^2Z^3\\
Q=&Z^1Z^2-Z^0Z^3\ .
\end{aligned} 
\end{equation}
From eq.~(\ref{twistor}), we then find
\begin{equation}
\frac{\partial}{\partial \omega^A}P=0\quad \text{and}\quad  
\frac{\partial}{\partial \omega^A}\frac{\partial}{\partial \omega^B} Q=0,
\label{conditions}
\end{equation}
so that it is straightforward to compute
\begin{equation}
\label{ASD_YM}
\tilde{f}^{\rm EM}_{AB}=-\frac{Q_A Q_B}{Q^2},\quad 
Q_A\equiv \frac{\partial}{\partial \omega^A} Q.
\end{equation}
The anti-self-dual Schwarzschild / Taub-NUT solution can be obtained
from the following twistor function for use in
eq.~(\ref{Penrose2})~\cite{Sparling}:
\begin{equation}
\label{Schwarzschild_ASD}
\tilde{f}^{\text{grav}}= \tfrac{1}{2}Q\text{log}\frac{Q}{P}
\end{equation}
with $Q$ and $P$ as defined in \eqref{P_and_Q}. Then
\begin{equation}
\label{ASD_grav}
\tilde{f}^{\text{grav}}_{ABCD} = \frac{Q_A Q_B Q_C Q_D}{Q^3} 
\end{equation}
Finally, comparing equations \eqref{ASD_YM} and \eqref{ASD_grav}, we
see that the double copy formula \eqref{ASD_DC} is indeed verified,
with
\begin{equation}
f=Q^{-1} 
\end{equation}
We remark that the expression for the scalar twistor function is the
same as that used for the self-dual analysis of
section~\ref{sec:Schwarzschild}, as must be the case. Furthermore,
different choices of the quadratic form $Q$ (subject to the conditions
of eq.~(\ref{conditions})) will map out the space of type D vacuum
solutions~\cite{Haslehurst}. \\

It is possible to extend the above to general families of solutions. Firstly, recalling the definition of $Z^\alpha$ eqn. \eqref{twistor2}
\be
Z^\alpha=(Z^0,Z^1,Z^2,Z^3)=(\omega^0,\omega^1,\pi_{0'},\pi_{1'})=(\omega^A,\pi_{A'}) 
\ee
we notice that 
\be
Q=Q_{\alpha\beta}Z^\alpha Z^\beta=Z^1Z^2-Z^0Z^3 
\ee
was chosen exactly such that
\be
\label{Qpi}
Q_A = \frac{\partial}{\partial \omega^A} Q =(-\pi_{1'},\pi_{0'})=(\epsilon\pi)_A,
\ee
so that $Q_A$ is just $\pi_{A'}$ rotated by a Levi-Civita symbol. It is then straightforward to show that we can write
\be 
\label{rhoxQ}
\rho_x[Q]=(\mathcal{A}^{A'}\pi_{A'})(\mathcal{B}^{B'}\pi_{B'})
\ee
with $\mathcal{A}^{A'}$ and $\mathcal{B}^{B'}$ defined as in \eqref{ABcaldef2}.
Then, using \eqref{Qpi} and \eqref{rhoxQ} we can write the integrand for the Coulomb solution of \eqref{FEMcoulomb} as:
\be 
\begin{aligned}
\rho_x\left[
\frac{\partial}{\partial \omega^A} \frac{\partial}{\partial \omega^B} \tilde{f}^{EM}\right]
=&-(\epsilon\pi)_A(\epsilon\pi)_B \frac{1}{(\mathcal{A}^{A'}\pi_{A'})^2(\mathcal{B}^{B'}\pi_{B'})^2}\\
=&-(\epsilon\pi)_A(\epsilon\pi)_B f^{(1,1)}_{EM}
\end{aligned}
\ee with $f^{(1,1)}_{EM}$ the function appearing in the self-dual
transform (see eq.~\eqref{f12EMdef}). Similarly, for the Schwarzschild
solution of eq.~\eqref{Schwarzschild_ASD} we have \be
\begin{aligned}
\rho_x\left[
\frac{\partial}{\partial \omega^A} \frac{\partial}{\partial \omega^B} 
\frac{\partial}{\partial \omega^C} \frac{\partial}{\partial \omega^D} \tilde{f}^{\text{grav}}\right]
=&2(\epsilon\pi)_A(\epsilon\pi)_B
(\epsilon\pi)_C(\epsilon\pi)_D \frac{1}{(\mathcal{A}^{A'}\pi_{A'})^3(\mathcal{B}^{B'}\pi_{B'})^3}\\
=&2(\epsilon\pi)_A(\epsilon\pi)_B (\epsilon\pi)_C(\epsilon\pi)_D f^{\rm TypeD}_{\rm grav.}.
\end{aligned}
\ee

Above we described the double copy for Type D. In order to progress to more general families, we will first use the results above to find the $\tilde{f}^{\text{EM}}$'s which map to $f^{(0,2)}_{EM}$ and $f^{(2,0)}_{EM}$ defined in \eqref{f12EMdef} and \eqref{f3def}. Making the ansatz
\be
\tilde{f}^{(0,2)}_{EM}=\frac{Q}{R^2}\text{log}\left(\frac{Q}{P}\right) 
\ee  
with $Q$ and $P$ as before and
\be
R=R_\alpha Z^\alpha=R^{A'}\pi_{A'}, 
\ee
we have
\be 
\frac{\partial}{\partial \omega^A} \frac{\partial}{\partial \omega^B} \tilde{f}^{(0,2)}_{EM}
=\frac{Q_A Q_B}{R^2 Q}.
\ee
Then
\be
\begin{aligned}
\rho_x\left[ \frac{\partial}{\partial \omega^A} \frac{\partial}{\partial \omega^B} \tilde{f}^{(0,2)}_{EM}\right]&=
(\epsilon\pi)_A(\epsilon\pi)_B \frac{1}{(\mathcal{A}^{A'}\pi_{A'})(\mathcal{B}^{B'}\pi_{B'})(R^{A'}\pi_{A'})^2}\\
&\xrightarrow[]{R^{A'}=\mathcal{B}^{A'}}(\epsilon\pi)_A(\epsilon\pi)_B f^{(0,2)}_{EM}.
\end{aligned} 
\ee
Similarly, we have
\be
\tilde{f}^{(2,0)}_{EM}=\frac{Q}{S^2}\text{log}\left(\frac{Q}{P}\right) 
\ee  
with $Q$ and $P$ as before and
\be
S=S_\alpha Z^\alpha=S^{A'}\pi_{A'}, \qquad S^{A'}=\mathcal{A}^{A'}. 
\ee
Finally, the anti-self-dual analogue of the gravity function \eqref{WeylType1} will be
\be
\tilde{f}_{\rm grav.}^{\rm (I)}=\left[\mathcal{G}_1 Q+ \mathcal{G}_2 \frac{Q^2}{R^2} +\mathcal{G}_3 \frac{Q^2}{S^2} +\mathcal{G}_4 \frac{Q^3}{S^2 R^2}\right] \text{log}\left(\frac{Q}{P}\right) 
\ee
with $P,Q,R,S$ defined as before. If we choose
\be
\mathcal{G}_1=-\tfrac{1}{2}\tfrac{[\mathcal{C}\mathcal{A}][\mathcal{C}\mathcal{B}]}{[\mathcal{A}\mathcal{B}]^2},\quad
\mathcal{G}_2=\tfrac{1}{2}\tfrac{[\mathcal{C}\mathcal{B}]^2}{[\mathcal{A}\mathcal{B}]^2},\quad
\mathcal{G}_3= \tfrac{1}{2}\tfrac{[\mathcal{C}\mathcal{A}]^2}{[\mathcal{A}\mathcal{B}]^2},\quad
\mathcal{G}_4=-\tfrac{1}{6}\tfrac{[\mathcal{C}\mathcal{B}][\mathcal{C}\mathcal{A}]}{[\mathcal{A}\mathcal{B}]^2}
\ee
then the double copy factorisation proceeds by direct analogy to \eqref{Weyltype1b} and the subsequent discussion.\\

The double copy formula of eq.~(\ref{ASD_DC}) is perhaps less
desirable than the form based on dual twistors, in that it ceases to
be a simple product, and thus appears to offer no additional
advantages with respect to the spacetime double copy formalism. Note
also that the same objections regarding how to interpret the procedure
in cohomological terms apply here. The twistor ``functions'' to be
entered into eq.~(\ref{Penrose2}) are actually cohomology classes,
which in this case are elements of the sheaf cohomology group ${\rm
  H}^1(\mathbb{PT},{\cal O}(n-2))$, for a spin $n$
field. Differentiating $2n$ times maps each cohomology class into an
element of ${\rm H}^1(\mathbb{PT},{\cal O}(-n-2))$, similar to the
case of self-dual fields. Once again, we may take the pragmatic view
that in order to generate a particular spacetime double copy, it is
sufficient to show that particular representatives of the cohomology
classes may be found in twistor space, that achieve the desired
spacetime relationship.

\section{Conclusion}
\label{sec:conclude}

In this paper, we have examined the Weyl double copy that relates
solutions of biadjoint scalar, gauge and gravity theories, using a
twistor-space formalism initiated in ref.~\cite{White:2020sfn}. The
latter argues that each instance of the Weyl double copy in spacetime
can be associated with a certain product of functions in
twistor space. We have provided full details of how this formalism is
sufficient to derive the previously noted form and scope of the Weyl
double copy, namely the fact that it applies to arbitrary vacuum type
D solutions. We have also gone further than ref.~\cite{White:2020sfn}
in providing examples of Petrov type I and II solutions in gravity, in
addition to types III, D and N. However, such solutions are limited to
linearised level, which is ultimately due to the limitations of the
Penrose transform itself. We have also shown how similar arguments can
be used to derive spacetime double copy formulae for anti-self-dual
fields, as well as self-dual ones.\\

Care must be taken in how to interpret the twistor space double copy,
given that it apparently involves multiplying together twistor
functions. In the Penrose transform, the ``functions'' are in fact
cohomology classes (i.e. elements of sheaf cohomology
groups). Deriving a given instance of the Weyl double copy then
amounts to showing the existence of appropriate representations of
each cohomology class, such that the functions entering a particular
instance of the spacetime Weyl double copy are indeed related by a
twistor-space product. This is a far cry from demanding a map between
the relevant cohomology groups themselves, and the investigation of
whether a more rigorous twistor-space interpretation exists deserves
further investigation, as it may shed further light on the ultimate
origins and scope of the double copy itself. It may also open up the
possibility to look at fully non-linear solutions. Work on these
issues is in progress.


\section*{Acknowledgments}

We are extremely grateful to Tim Adamo for illuminating conversations,
and comments on the manuscript. We also wish to thank Andreas
Brandhuber, Andr\'{e}s Luna, Gabriele Travaglini and Costis
Papageorgakis for discussions. This work has been supported by the UK
Science and Technology Facilities Council (STFC) Consolidated Grant
ST/P000754/1 ``String theory, gauge theory and duality'', and by the
European Union Horizon 2020 research and innovation programme under
the Marie Sk\l{}odowska-Curie grant agreement No. 764850 ``SAGEX''. EC
is supported by the National Council of Science and Technology
(CONACYT). SN is supported by STFC grant ST/T000686/1.

\bibliography{refs}
\end{document}